\documentclass[prb,twocolumn,showpacs,preprintnumbers,amsmath,amssymb]{revtex4}

\usepackage{graphicx}
\usepackage{dcolumn}
\usepackage{bm}

\begin{document}

\preprint{APS/123-QED}

\title{Processes in Resonant Domains of Metal Nanoparticle Aggregates\\ and Optical Nonlinearity of Aggregates in Pulsed Laser Fields}

\author{Anatoliy P. Gavrilyuk}
 \affiliation{Institute of Computational Modeling of Russian Academy of Sciences,  Siberian Branch, 660036, Krasnoyarsk.}

 \email{glyukonat@icm.krasn.ru}

\author{Sergei V. Karpov}
\affiliation{
L. V. Kirenskiy Institute of Physics of Russian Academy of Sciences, Siberian Branch, 660036, Krasnoyarsk.
}
 \email{karpov@iph.krasn.ru}
\date{\today}

\begin{abstract}
The specific optical nonlinearities inherent in aggregates of metal nanoparticles under pico- and nanosecond pulsed laser irradiation are studied in nanoparticle aggregates formed in silver hydrosols. The results of experimental studies of the correlation between the degree of aggregation of silver hydrosols and their nonlinear refraction index ($n_{2}$) at the wavelengths 0.532   and 1.064  $\mu$m are discussed. The experiments revealed that $n_{2}$ changes its sign at 1.064 $\mu$m  as the degree of the hydrosol aggregation grows. The role of various processes occurring in resonant domains of aggregates and the kinetics of these processes under laser irradiation resulting in dynamic variation of the polarizability of aggregates are analyzed. The areas under study included the kinetics of particles displacement considering dissipative forces, heating of the particles and of the surrounding medium depending on the wavelength, intensity and duration of laser pulses. A theory of interaction of laser radiation with an elementary type domain - two bound silver nanoparticles - was developed to describe the kinetics of resonant domains photomodification in aggregates. This theory takes into account thermal, elastic, electrostatic and light induced effects. The experimental results on laser photomodification of silver particle aggregates are discussed in the context of our model. These results include photochromic and nonlinear optical effects, in particular, the nonlinear refraction and nonlinear absorption.
\end{abstract}


\maketitle

\section{\label{part1}Introduction}

Nonlinear optical media consisting of aggregates of metal nanoparticles have attracted significant attention over the past decade due to their unique properties and high application potential (see, in particular \cite{ShalPRep,Shalbook,bookKarpov,stockman} ).

Primary examples of such media are sols containing fractal aggregates of colloid silver nanospheres and random metal-dielectric nanocomposites \cite{ShalPRep,Shalbook,bookKarpov}.
Resonant excitation of surface plasmons in such systems results in a giant enhancement of local electromagnetic fields  \cite{ShalPRep,Shalbook}. This enhancement was observed in the spectral range where the linear absorption spectra of aggregates were inhomogeneously broadened while the quality factor of the surface plasmon modes was found to increase with the wavelength.

New physical models proposed in 1980-90's (see, e.g.  \cite{ZP88,stockmanMarkel,OAO,stPOlMarkel}),
predicting various nontrivial optical properties of these materials gave an impetus to further investigations. Already the early experiments with silver hydrosols containing fractal aggregates proved that these materials feature unique nonlinear optical properties \cite{but} (also see the review in  \cite{ShalPRep,Shalbook,bookKarpov}.

It was found in \cite{but}
that the intensity of the four-wave degenerate mixing is enhanced by five to six orders of magnitude in aggregated silver hydrosols as compared to nonaggregated hydrosols. Other nonlinear optical effects experimentally discovered in aggregates of silver and gold nanoparticles include nonlinear optical activity (nonlinear gyrotropy)
\cite{Drach},  spontaneous rotation of the polarization ellipse~\cite{Drach}, nonlinear absorption and refraction
\cite{93,Drach,Lepeshkin,94,95},
the reverse Faraday
effect~\cite{93,Drach,Lepeshkin,94,95},
optical Kerr effect~\cite{Lepeshkin,QE01}, harmonic generation~\cite{Ganeev},
enhancement of nonlinear responses of organic molecules adsorbed on
nanoparticle surfaces~\cite{Zhur} and photochrome reactions
that mediate the effect of optical
memory~\cite{JETP88,PRL98,JTP03,Ind,Saf96,104}.

A considerable wealth of knowledge about nonlinear optical effects in random heterogeneous nanostructures has been accumulated by now. However, there is still a lack of a comprehensive understanding of the origin and the mechanisms of various nonlinear optical effects, especially for fast nonlinearities with the characteristic times of the order of
$10^{-12}{\rm }$ s. It should be noted that nonlinearities with the  times from
 $10^{-14}{\rm}$ s to $10^{-12}{\rm}$ s due to optical
excitation of electrons in individual nanoparticles were discussed in
Refs.~\cite{Flyz,Raut}. As for the mechanisms of slow nonlinearity (with the characteristic times of the order of $10^{-7}{\rm}$ s), these are usually associated with thermal expansion and electrostriction and are outside the scope of our discussion.

One of the sources of fast nonlinearity is the effect of spectrum- and polarization-selective transparency in aggregated sols induced by strong laser pulses. This effect was discovered in Ref.  \cite{JETP88}  and studied in detail in Refs.~\cite{but,PRL98,JTP03,Saf96}.
It happens when the laser frequency falls within the spectral range of resonant surface plasmon excitation of aggregates and the laser power exceeds a certain threshold. The effect is attributed to the local photomodification of aggregates and has the characteristic times from  $60{\rm} $ ps to
$150{\rm} $ ps \cite{bookKarpov,but,JTP03,Ind} depending on the irradiation intensity. Photomodification of the nanoaggregate structure takes place at laser intensities going above a certain threshold. However, the photomodification is confined to relatively small resonance domains \cite{PRB05,JCP06,CJ07} and does not affect the large scale geometry of the aggregate.

The optical memory effect is associated with laser photomodification of nanoaggregates. It is believed that the structure of resonant domains of aggregates is subject to selective changes during this process  \cite{Saf96}. The changes take place both in the size, shape and state of the resonant particles (their fusion or evaporation) as well as in their relative arrangements. Thus, for example, in Ref. \cite{QE01}
it is suggested that we deal here with dynamic plasmon resonances of silver nanoparticles whose frequency changes under the applied pulsed radiation. According to \cite{QE01}, this is caused by evaporation of particles in the laser pulsed field and the formation of a rapidly expanding plasma inside which the electron concentration is decreased. In
\cite{Drach,Claro1,Claro2,Claro3,Hallok} a possibility of particles displacement in aggregates is shown under the action of light induced dipole (generally - multipole) forces. The collapsed adsorption layer of particles during photo- and thermo-emission of electrons from the metal core was shown to result in the same effect as in photostimulated aggregation of metal sols \cite{CJ2002}.

In order to analyze the regularities revealed in nonlinear refraction and nonlinear absorption of metal hydrosols, one should consider in detail the whole set of processes of interaction of aggregate particles with the external radiation, including the processes occurring at the interface. These regularities include, in particular, the change of the nonlinear refraction sign with the growing degree of aggregation of silver hydrosol \cite{QE01,Ganeev} at 1.064 $\mu$m.
However at present the physics of photomodification of fractal nanoparticle aggregates with plasmon absorption has not been studied in enough detail yet and a theory that could be used to describe the kinetics of photomodification and to determine the role of the processes underlying this effect is not available either. Such a theory would help to unriddle some of the perplexing issues concerning the nature of photomodification and the nonlinear optical phenomena observed in sols with fractal aggregates under applied laser fields.

The aim of this paper is to study the mechanisms of fast (with times of order  $10^{-8}-10^{-12}$~s) nonlinear-optical responses of nanoparticle aggregates in metal sols (above all, in the sols with a liquid surrounding medium) induced by the optical Kerr effect which can help to explain the features of nonlinear optical processes in metal sols.
\section{\label{part2}Experimental results and discussion}
Silver hydrosols are the most suitable model medium for studying optical phenomena in metal colloids. Their optical properties in the visible range are determined by an isolated surface plasmon resonance lying outside the interband absorbtion range. Coagulation of particles in the process of their Brownian motion leads to formation of colloid structures with a fractal distribution of particles (see Refs. in
\cite{ShalPRep,Shalbook,bookKarpov}).

The growth of fractal aggregates results in a strong inhomogeneous broadening of the plasmon absorption spectrum of hydrosols due to a strong electrodynamic interaction between the particles
 \cite{Shalbook,bookKarpov}.
Quite a number of experimental and theoretical papers  \cite{Shalbook,bookKarpov,PRB04} have proven by now the existence of a correlation between the degree of aggregation of colloids and their optical properties.

The simplest preparation technique of a standard $Ag$  hydrosol uses collargol (a silver antiseptic), containing disperse silver (the methods of preparation of sols are described, in particular, in
\cite{but,Heard}). The sizes of silver nanoaggregates in this type of sols are within the range  100-1000 nm, the average size of silver particles being  4 to 20 nm and a volume fraction of silver being of the order of  $(1\div5)\cdot10^{-6}$.

 Particles of this type of hydrosol feature a water soluble polymeric adsorption layers (adlayer) which prevents their fast coagulation and direct contact of particles in aggregates, as well as enhances the sedimentation stability.

In the used type of hydrosol the adlayers consisted of casein, the adlayer thickness being 10-20\% of the particle radius. In the absence of an adlayer, a hydrosol with the given concentration of the disperse phase would aggregate within less than
$10^{-2}$~s followed by fast sedimentation.
The thickness of the adlayer particles appears to be several times smaller in aggregates than in isolated particles  \cite{bookKarpov,CJ2002}. In the case of collargol, the thickness of the adlayer was reduced to accelerate the aggregation process by adding  $NaOH$  (0.1 mol/l) into the solution, which leads to dehydration of the casein shell and reduces its volume.

\begin{figure}
\includegraphics{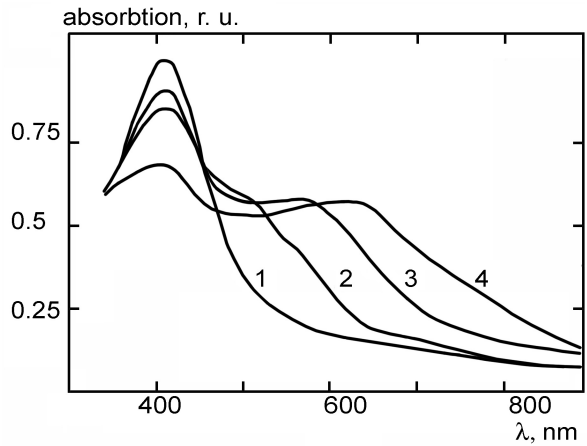}
\caption{\label{fig1}Typical absorption spectra of collargol-based silver hydrosols for various degrees of aggregation
 $A$: Curve 1 is a nonaggregated sol, Curves 2-4 show the hydrosol at different stages of aggregation. Curve 4 corresponds to the developed stage of hydrosol aggregation. The absorption spectra of silver hydrosols differ from each other at various stages of aggregation, in particular: 1 - $A$=0 (isolated particles), 2 - $A$=0.2, 3 - $A$=0.8 (intermediate stages of aggregation), 3 - $A$=1 (all particles are incorporated in fractal aggregates).}
\end{figure}

The degree of aggregation ($A$) was monitored by the absorption spectra of hydrosols. Measuring of the absorption spectrum broadening to determine the degree of aggregation was suggested, in particular, in \cite{bookKarpov,Ind} (where $A$ is the normalized broadening of the spectrum in the range
$0 \leqslant A\leqslant 1$).  Actually, the proposed parameter, $A$,  is defined as the difference between specific integrals over the functions describing one of the spectral curves  at the stage of aggregation ($2-4$ in  Fig.~\ref{fig1}) and the original Curve  $1$ in the wavelength range from approximately $\lambda_1>440$~nm (the first point of intersection of the curves) to the long-wavelength point of intersection
$\lambda_2$ ($A\propto
^{\lambda_2}_{\lambda_1}E^{abs}_i(\lambda)d\lambda$
$-\int^{\lambda_2}_{\lambda_1}E^{abs}_1(\lambda)d\lambda$),
 $E^{abs}_i(\lambda)$ is the spectral form-factor (contour) of the i-th absorption band. In particular,
 the $A$  value corresponding to Curve $3$  in Fig.~\ref{fig1} is equal to 0.8.
 In \cite{bookKarpov,Ind} it was shown that  the parameter $A$  correlates with the rate of coagulation, i.e. $A$  is proportionate to
$N_0(t)^{-1}$  (where $N_0(t)$   is the full number of particles in the system including many-particle ones) over the time of double reduction of the initial ($N$) concentration $N_0(t)=N/2$.

In most cases, single beam longitudinal scanning (along the designated $Z$ axis) \cite{Zscan} is a convenient method to study nonlinear optical proiperties of materials. The advantage of this method is in its simplicity and high sensitivity of measuring nonlinear refraction and absorption indices. We used a $Q$-switched $Nd:YAG$ laser in our experiments. The behaviour of the nonlinear refractive index  $n_{2}$  was studied at the laser wavelength $\lambda $=1.064 $\mu$m ($W$=12mJ, $\tau $=15 ns).

The experimental setup was built using the arrangement with a limiting aperture \cite{QE01,Ganeev,Zscan}.
A 1 mm diameter aperture, which transmitted $\sim $1{\%} of laser radiation, was placed at a distance of 100 cm from the focal area.
Behind the aperture, a photodiode was placed whose output signal was measured. To avoid the influence of instability of the laser output on the measurement result, the signal detected with the photodiode was normalized to the indications from another diode placed in front of the focusing lens. The laser radiation was focused with an 18-cm focal length lens.

A 2-mm thick quartz cell containing silver hydrosol travelled along the optical axis $Z$ driven by a micropositioner, passing the areas in front and behind the focus through the focus. The focal spot size at the beam waist was 100 $\mu$m and the highest radiation intensity achieved was
8$\times $10$^{9} $ \  W/cm$^{2}$. The arrangement with a limiting aperture enabled us to measure both the sign and the absolute value of the nonlinear refractive index   $n_{2}$  of the medium under study.

\begin{figure}
\includegraphics{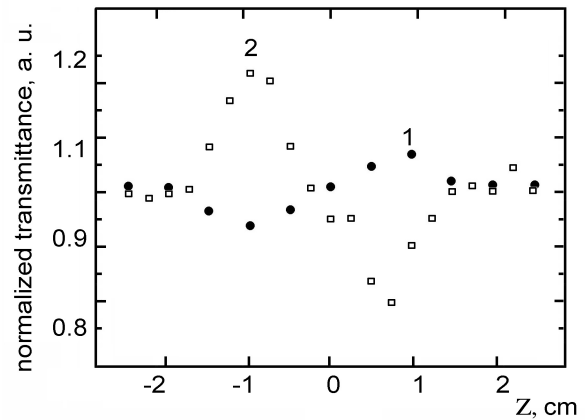}
\caption{\label{fig2}Normalized transmission vs $Ag$ hydrosol cell position in the limiting diaphragm arrangement. Curve 1 was obtained at the degree of aggregation $A$=0, Curve 2 - $A$=1.}
\end{figure}

\begin{figure}
\includegraphics{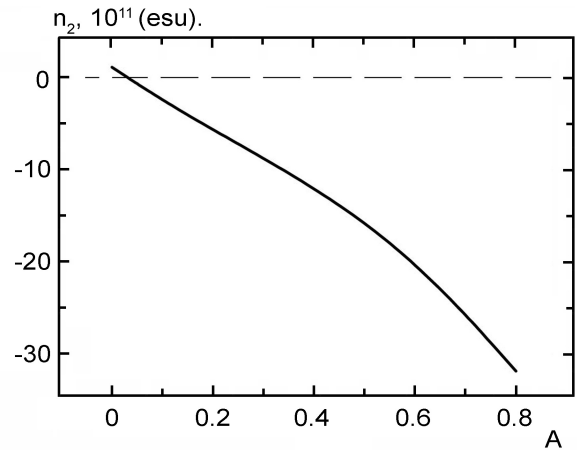}
\caption{\label{fig3}Nonlinear refractive index vs the degree of aggregation.}
\end{figure}

Fig.~\ref{fig2} shows typical dependencies  $T(Z)$  of the $Ag$  hydrosol cell transmission on the cell position with respect to the laser radiation focus.
 These dependencies were obtained for the initial (Curve 1:  $A$=0, the intensity in the focal area is  $I\simeq 10^9 $ W/cm$^2$) and the final ($A$=1,  $I\simeq 10^8$ W/cm$^2$) stages of aggregation. It has been found that the sign of  $n_2$  changes with the increasing $A$ \cite{QE01}. Similar results were obtained in aggregated silver hydrosols
 when irradiated by single picosecond pulses
 $\tau$=35-50 ps   \cite{Ganeev}.

 Nonlinear optical parameters of the medium were determined from optical transmission
using the $Z$-scan equation \cite{Zscan}
\begin{equation}\label{1}
{\Delta T}_{p-v}{=0.404(1-S)}^{0.25}{ \mid2\pi \Delta n_0}{
[1-\exp(-\alpha L)]/(\lambda \alpha )\mid},
\end{equation}
\noindent
where ${\Delta T}_{p - v}$ is the normalized difference between the maximum and minimum transmission
in the dependence $T(Z)$, $\Delta
n_{0}$ is the nonlinear addition to the refractive index ($\Delta n$)
at the focus ($Z=0$), $S$  is the geometrical transmission of the aperture
(a fraction of radiation in a diverging Gaussian beam incident on the photodetector
through the aperture  in the absence of nonlinear effects),  $\lambda $
is the radiation wavelength; $L$ is the sample length, $\alpha $ is the linear absorption coefficient.

For cubic nonlinearity, the refraction index  $n$  is related to $n_2$ through the known equation
 \cite{Shen,Zscan} $n =
n_{0} + 0.5n_{2} {\left| {E} \right|^{2}} = n_{0} + \Delta n$, where $n_{0}$ is linear
 refractive index, $E$  is the  amplitude of field intensity.
 The nonlinear refractive index   of silver hydrosol calculated from Equation (\ref{1})

 An important experimental result obtained in this paper, which should be
discussed separately, is the change of the sign of $n_{2}$ with increasing $A$ (Fig.~\ref{fig3}).
In fact, the initial ability of the medium for self-focusing transforms into its ability for self-defocusing with the increasing  $A$.
Explanation of this effect gets our primary attention in this paper because it gives us an instrument to estimate the accuracy of our model and whether this model can be used for explanation of other nonlinear phenomena. The first attempt to explain the effect was made in \cite{QE01}; some of the important questions however (see below) were left unanswered. In our paper as well as in \cite{QE01} it is assumed that nonlinear refraction in Ag hydrosols is caused by the optical Kerr effect at resonances existing in the medium, because, as follows from our estimates, the contribution from thermal expansion of the medium is negligible in the case of nanosecond pulses used in the experiments.

The nonlinear refractive index of a medium is related to the third-order nonlinear susceptibility as   $n_{2} \propto n_{0}^{ - 2} Re[\chi ^{(3)}]$ \cite{Zscan}.
The Kerr nonlinear susceptibility (its real part) $Re[\chi ^{(3)}]$  of a multicomponent system for the process $\omega = \omega + \omega - \omega $  in the case of a simplest two-level system away from resonances can be described as (see, for example, (see, for example, \cite{Shen})
\begin{equation}\label{2}
Re[\chi ^{(3)}(\omega , - \omega ,\omega )] \propto \sum N_i
{(d_{12} )_i ^4}{[(\omega _{12})_i - \omega ]^{-3}},
\end{equation}
\noindent where $(d_{12})_{i}$  is the matrix  element of the electric dipole moment at the frequency
$(\omega _{12} )_i$, $N_i$ is the concentration of particles of the $i$th component of the medium (summation
is performed over all components). One can see from
this expression  that the sign of $n_2$ is determined by the sign of the resonance detuning.
According to experimental data the nonlinear refractive index of water is close to zero. Nevertheless it should be noted that nonlinear dispersion properties of water can change under the influence of nanoparticles.

In out paper as well as in \cite{QE01} we assume that nonlinear refraction in Ag hydrosols
is caused by the optical Kerr effect at the resonances existing
in the medium, because, as follows from our estimates, the contribution
from thermal expansion of the medium in the case of nanosecond pulses
used in experiments is negligible.

The fact of changing of $n_{2}$ for hydrosols upon their aggregation is known; however,
the models describing this effect do not predict the sign change (at least,
upon irradiation at 1.064 $\mu$m). Therefore, this effect should be additionally analyzed.

According to experimental data nonlinear refractive index of a water at 1.064 $\mu$m is close to zero.
Nevertheless we note that nonlinear dispersion properties of water  can change under the influence
of nanoparticles.

The primary idea of this paper is based on the assumption that the local structure of nanoaggregates in the resonant domain area changes under the applied laser field due to the changes in the distances between adjacent particles. We shall call this change as "photomodification" in our further discussion. The change of the local structure causes changes in electrodynamic interparticle interactions, which affects the plasmon absorption spectra to result in the change of the denominator sign in Expression (\ref{fig2}). Such local reconstruction can be associated with the shift of particles both under the influence of light induced multipole interaction and due to the collapse of adlayers of the particles, as shown in \cite{CJ2002}, and it is believed to be one of the reasons for photostimulated aggregation of metal sols. What would be the consequences of irradiation of nanoaggregates by high intensity laser light?

\subsection{\label{part2_1}Optical Kerr effect in the case when photomodification of aggregates is negligible}

Employing Expression (\ref{fig2}) for qualitative description, one can see that the sign of $n_{2}$ depends on the resonant detuning sign. If the condition $\omega <\omega_{12}$
 is fulfilled, self-focusing is observed in
a one-component medium, while for $\omega
>\omega _{12}$ self-defocusing is observed.
If upon summation over components of the medium the partial contribution
of some of the components dominates, the sign of $n_{2}$ will also depend
on $(d_{12})_i$ and the concentration of the $i$th component. It is quite possible that the dominating role of one of the resonances in the alteration of nonlinear dispersion properties of the medium and interference of the Kerr nonlinear polarizations at certain resonances are the factors resulting in a positive or negative nonlinear refraction.
 In our case, silver nanoparticles and water were the main components of the medium.

Consider contribution from each of the main components of hydrosol
into $Re[\chi^{(3)}]$ at the laser wavelength 1.064 $\mu$m taking into account
the resonances  typical for such kind of media. In the case of silver particles,
a surface plasmon with the absorption band in the range  $\lambda _{pl}$=0.4-0.42 $\mu$m and
the interband absorption band lying in the range   $\lambda <0.35$ $\mu$m play the main role
in the formation of a nonlinear response. The frequencies of both these bands
are higher than the laser frequency ($\omega _{l})$ hence their contribution
to nonlinear refraction will be positive.
The inhomogeneous broadening of the  surface plasmon spectrum
along with the effective red shift of the absorption band (Fig.~\ref{fig2})
during aggregation of silver particles results in the increase
of the positive contribution to nonlinear refraction.

However just the aggregation  of $Ag$ particles
on its own should not result in the change of the nonlinear refraction  sign
because even in the case of maximum spectral broadening and partial absorption
in the region $\omega <\omega _{l}$, the weight contribution to $\chi ^{(3)}$  of
the resonances located in the region $\chi ^{(3)}$ prevails.

The Kerr susceptibility $Re[\chi ^{(3)}]$ of water at $\lambda=1.064 \
\mu$m is associated with the influence of
the electron-vibronic band $\lambda <0.19$ $\mu$m)  having the integrated ($d_{12} )_{e - v}$ 10$^{-18}$ esu
 and of the vibronic band having the fundamental transition at ($\lambda \approx
$2.9 $\mu$m)
and its first overtone at ($\lambda \approx $1.45 $\mu
$m) \cite{115,116}.

The contributions from these bands to $\chi ^{(3)}$  have  opposite signs,
and the net nonlinear refraction will be determined by the
 ($d_{12} )_{e - v}$ to ($d_{12} )_{v}$ ration. Our estimates show that the contribution from
the short-wavelength transitions dominates, and nonlinear refraction of
water at $\lambda = 1.064$ $\mu$m is positive, which agrees with the experimental data \cite{13,14}.

This is due to $(d_{12} )_{e - v}^{4} \vert
\Delta \omega _{e - v} \vert ^{ - 3}$  somewhat higher than $(d_{12})_v^4
\vert \Delta \omega _v \vert ^{ - 3}$ ($\Delta \omega _{e - v}$  and
$\Delta \omega _{v}$
 are the detunings of the respective resonances from $\omega _{l}$).

That  is, the contribution from the both medium components to $Re[\chi ^{(3)}]$
is positive and should increase during  aggregation of particles.
 This however contradicts the experimental data and requires the model to be adjusted.

The thermal mechanism underlying vibrational excitation of water molecules, which are in contact with strongly heated silver particles or with the resultant plasma, is another probable cause of changes occurring in the hydrosol nonlinear dispersive properties. This leads to the population of high vibrational states of water molecules ($v\gg 1$), the electric dipole moment between which, as known, increases as
$(d_{12} )_{v}\propto \sqrt{v}$ \cite{Eljashevich} (where $v$  is the number of the vibrational level). A similar mechanism of self-focusing due to vibrational rotational, and electronic excitation of molecules and atoms was discussed in Ref. \cite{16}.

 The estimated minimal value of unperturbed value $(d_{12})_{pl}$  based
 on spectroscopic data for $Ag$ sol with the particle  size  $ \sim $10 nm  is  2.5$\times
$10$^{ - 16} $  esu.
 To estimate ($d_{12} )_{v}$ for the vibrational transition of water ($v$=0-1) nearest to  $\omega _{l}$
one can use the values of the  optical constants of water at this frequency \cite{115,116} and
 the Lorenz-Lorentz formula. The calculation yields   $(d_{12} )_{v} \sim $10$^{ - 20} $ esu.

With the above values of $(d_{12})_{pl}$   and ($d_{12} )_{v}$   and the ratio $\mid\Delta \omega _{pl}^3 / \Delta \omega
_v^3 \mid \approx 10^4$,  and for the volume fraction of silver ($N_{Ag}
\leqslant 10^{ - 6})$, the ratio of nonlinear polarizations at the transitions in silver and in water for nonaggregated sols,
\begin{equation}\label{3a}
 P_{Ag}^{(3)} :P_{H_2 O}^{(3)} \propto [N_{Ag} (d_{12}
)_{pl} ^4\vert \Delta \omega _{pl} \vert ^{ - 3}]:[(d_{12} )_v^4
\vert \Delta \omega _v \vert ^{ - 3}]
\end{equation}
will be equal to 10$^7$.

The experimental data show that the value of $\chi^{(3)}$ for hydrosol is several orders
of magnitude higher as compared to pure water, although it is underestimated
taking into account the large value of $(d_{12})_{pl}$.
This suggests that silver plays an important interceding role in the enhancement
of optical nonlinearity of hydrosol.

Consider some examples describing the conditions for manifestation of this mechanism.
The dissociation energy of
a water molecule is close to the energy of the upper vibrational state and
is 8.4$\times $10$^{ - 20}$ J (see Ref. \cite{17}), which corresponds to the temperature 6$\times $10$^4$ K.
The energy $ \sim
$10$^{ - 10}$J absorbed by a particle is sufficient for
thermal dissociation of 10$^8$  molecules, which occupy the volume 3$\times $10$^{ - 15}$ cm$^3$
and can be arranged around the particle in a spherical layer of thickness
$R$=10$^{ - 5}$ cm.

 The time during which all the molecules in the layer
will be dissociated is $ t=R/\rm V_s\sim 10^{ - 10} $ s, where $\rm V_s$  is the speed of sound in water.

This means that the volume occupied by thermally excited water molecules
is at least 10$^3$ times larger than the volume of solver particles, although
the amount $N_{t}$ of excited water molecules does not exceed $10^{ -
3}-10^{ - 2}$.

Assuming that $ (d_{12})_v$  increases by at least one order of magnitude,
and taking into account the ration $\vert \Delta
\omega _{e - v} \vert ^3/\vert \Delta \omega _v \vert ^3 \approx
10^6$ and
the above range of values of $N_{t}$, the contribution of vibrational
transitions of water into the Kerr susceptibility $\chi^{(3)}$  can become equal
to the contribution of the vibronic band,  substantially changing thereby Ratio (\ref{3a}).

\subsection{\label{part2_2}Contribution of thermal expansion of medium}

Laser photomodification of nanoaggregates as one of the factors accountable for
 the sign change of $n_2$  in nanosecond pulses should be analyzed further
 from the point of view of kinetics of  this process  when the experiment is carried out with picosecond pulses.
 Will the resonant domain  have time to change its structure (at moderated intensities)
 or to evaporate (at high intensities) within the  pulse duration
 $\tau $=30 ps?

To answer this question,
we will use the time dependencies ($t_{p})$ describing the development of
photomodification in such $Ag$ hydrosols \cite{Ind,but}.
It follows from \cite{Ind} that $t_p$ depends on the pulse energy density and decreases
from 150 to 50 ps upon the fourfold increase of the energy density.
Obviously, at higher energy densities, the time $t_p$ can further decrease,
and the particles can evaporate during one pulse, i.e. the role of this process
in the Kerr nonlinearity seems to be realistic from the point of view of kinetics for picosecond pulses.

Note, that the use of picosecond pulses in $Z$-scan technique
 led to similar: the change of a sign $n_2$  with the subsequent increase of
 its absolute value in the process of  aggregation of sols \cite{Ganeev}.

This lead us to important conclusion about the negligible contribution of the thermal component of nonlinearity caused by heating of the medium under applied radiation and a subsequent expansion of the heated region. Such heating reduces the medium density and hence decreases its refractive index as is the case with self-defocusing. The negligible contribution of the thermal mechanism to self-defocusing is confirmed by an insignificant expansion of the heated region even compared to the focal waist size during irradiation by a nanosecond pulse. Indeed, the size of the thermal lens being formed is $r_{0}=\rm V_s\times \tau $, and for  $\tau$ =15 ns  pulses it does not
exceed $ \sim 10^{ - 3} $ cm.

For 30 ps pulses $r_{0}$ will be only $\sim 10^{ - 6}d_b$, where $d_b$ is the beam diameter.
This is an order of magnitude smaller than the experimental size of the focal waist and at least two orders of magnitude smaller than the beam diameter outside the focus (1 cm away from the focus) where self-defocusing is maximumal (see. Fig.~\ref{fig2}).

When analyzing the conditions for manifestation of nonlinear refraction,
one should take also into account the scattering of laser radiation by metal
hydrosols related to the formation of vapor bubbles around nanoparticles
 absorbing radiation \cite{Chastov} (the time of their formation
and growth was  $10^{-8}$ s (see Ref. \cite{Chastov})).

However, when a cell with hydrosol is located
in front of the focus this effect will only prevent the Kerr self-defocusing.
This means that the Kerr nonlinearity on dynamic plasmon resonances of
silver and vibrational transitions in water makes a much greater contribution
than under real experimental conditions.

In estimating dynamic frequency  shifts one should also
also allow for the shift of  $\omega _{pl}$
due to the change in the  permittivity ($\varepsilon _{m}$)
of the environment when a vapor cavity is formed around the particle, because
$\omega _{pl} \sim 1/\sqrt{1+2\varepsilon _{m}}$ Ref. ~\cite{Boren}. This a correction does not exceed  ($\omega
_{pl})_{vapor}/(\omega _{pl})_{water} \leqslant $1.12.

Note that the above analysis did not take into account changes in the local structure of the nanoparticle aggregates themselves under modifying laser field. These can be done by modeling the processes and their kinetics occurring in resonant domains of large aggregates under the action of electromagnetic radiation. The following section is devoted to the formulation of such a model.
\section{\label{part3}Dynamic local restructuring  of fractal nanoparticle
        aggregates as a dominant source of optical nonlinearity and photochrome effects}

\subsection{\label{part3_1}External field model of resonant nanoaggregate domain}
Consider a model of interaction of a simplest aggregate - a bisphere consisting of two spherical  nanoparticles of equal size, with the external radiation. Besides, the bisphere (dimer) or its modification - trimer (consisting of three particles), can be treated as a component of a more complex aggregate in which it plays the role of a resonant domain. Such domains that are resonant to the given wavelength are randomly located at different sites of the large aggregate. Of course, such a domain is not electrodynamically isolated in a fractal aggregate, but the knowledge of its basic properties and their change under the influence of surrounding particles (eigen frequencies, relaxation constants, etc…) would help to predict the response of the fractal aggregate to the resonant laser radiation. To build such model, let us considered basic types of interactions in a bisphere (interaction of particles with each other and with the external optical radiation).

\subsubsection{\label{part3_1_1}Interaction with radiation}

We will limit our consideration to the discussion of  the interaction of optical radiation
 with the elementary model of a resonant bisphere consisting of two identical spherical silver nanoparticles (Fig.~\ref{fig1}). The dipole moment of such a bisphere $ {\bf P}$, generated by field  $\bf E$, is $ {\bf P}=\alpha _d \bf E$, the
polarisability of bisphere is $\alpha_d$ is represented by the sum of responses of single oscillators:
\begin{equation}\label{3}
\alpha_d(\omega) =\sum_j \frac{A_j}{\omega^2_{0j}-\omega
^2+i\omega \Gamma _j }, \quad\quad j=1, 2, ...,
\end{equation}
where $\omega_{0j}$ is resonant frequency of $j$th oscillator, $\Gamma _j$   is the relaxation constant,
$ A_j$ is coefficient. And  $\omega_{0j}$, \ $ \Gamma_j$, \ $ A_j$  are the functions
of a particle radius ($r_0$) and distances between surfaces of the particles included in bisphere ($h$).

\begin{figure}
\includegraphics{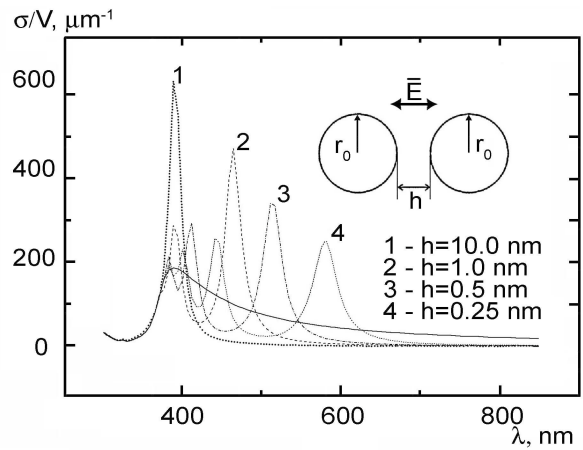}
\caption{\label{fig4}The absorption spectrum of $Ag$ bispheres in the longitudinal direction of polarization direction  and for various interparticle gap values $h$.  The inset shows the bisphere structure and the external field orientation (the bisphere particle radius is $r_0=10$ nm, $V$ is the volume of particles). Solid line corresponds to $h$=0 nm.}
\end{figure}

An absorption spectrum in the longitudinal direction of polarization has been calculated for $Ag$ bisphere particles with  $ r_0=10$  nm using the multipole approach \cite{PRB04} (with the account of up to the 32th order multipoles). The spectrum has been calculated for different values of the interparticle gap, $h$, and for the medium refractive index in the wavelength region under study being $n$=1.4 (Fig.~\ref{fig4}). The optical constants for silver were taken from \cite{Christi} and adjusted for the finite size-effect in small metallic spheres \cite{Boren, Markel}.

As seen in the Fig.~\ref{fig4}, the plasmon absorption spectrum  ($\lambda>400$ nm) is represented
by oscillating function. To find the analytical form of this function, which is necessary
in further calculations, it is convenient to use the approximated expressions
and to describe  spectral curve as the sum of quazi-lorentzian contours of two single-resonant
oscillators. The approximated analytical expressions for resonant frequencies of the long-wave and the short-wave oscillators (neglecting low-amplitude maxima in the wing of the bisphere absorption spectrum) are

\begin{equation}\label{4}
     \omega_{01}=\Omega _{01}\cdot 10^{15} s^{-1},
     \Omega _{01}=4.93 \exp \left (-\frac{0.252}{0.32+h}    \right ),
\end{equation}
\begin{equation}\label{4a}
     \omega_{02}=\Omega _{02}\cdot 10^{15} s^{-1},
     \Omega _{02}=4.93 \exp \left (-\frac{0.345}{0.72+h}    \right
     )+0.8,
\end{equation}
where $h$  is interparticle gap (in nm).

As particles move apart
 ($h\longrightarrow \infty$) \
$\omega_{01}$ approaches
the plasmon resonance frequency $\omega_{pl}$  of an isolated particle ($\lambda\sim 0.39 \mu$m) whereas once  $\omega_{02}$ reaches  $\omega_{pl}$ it stops changing and approximation by Expression (\ref{4a}) is no longer valid. However considering that the bispere interaction with the long-wavelength radiation ($\lambda >0.6$ $\mu$m) is more important in our model,
the absorption band with  $\omega_{02}$
in this case can be ignored.

The values of the amplitude factors   $A_{1}, A_{2}$ in (\ref{3}) were approximated by expressions:

\begin{equation}\label{5}
  A_1=\frac{10^4c}{4 \pi \varepsilon ^{1/2}_m} V \Gamma_1 \sigma '
  _1, \qquad
     A_2=\frac{10^4c}{4 \pi \varepsilon ^{1/2}_m} V \Gamma_2 \sigma '_2,
    \end{equation}
  \begin{equation}
  \sigma_1 ' =\frac{1.28 \cdot 10^2}{1,93+(\Omega_{01}-4.96)^2}
  (cm^{-1}),
  \qquad
   \sigma_2 ' =280 (cm^{-1}),
    \end{equation}
where $\sigma_1 '$ and   $\sigma_2 '$   are absorption cross-sections of radiation
in the unit  volume of a particle at  exact resonance (see Fig.~\ref{fig4}), $V$ is a particle volume,
$\varepsilon _m$ is the relative dielectric permittivity of medium (in the case of hydrosols it is a water),
$c$ is the  speed of light in vacuum.

Relaxation constants $\Gamma _1$  and
$\Gamma_2$  are determined by the expression \cite{Boren}:
\begin{equation}\label{6}
  \Gamma_1=\Gamma_2=\Gamma=\Gamma_b+\frac{V_F}{r_0},
\end{equation}
where $\Gamma_b=3\cdot10^{13} s^{-1}$ is the relaxation constant of free electrons in a
 metal bulk (in our case $Ag$), $V_F$ is  the  Fermi velocity of electrons, $r_0$ is a particle radius.

Knowing the bisphere  polarisability $\alpha_d$  we can determine the absorption cross-section for laser radiation
at the  frequency:
\begin{equation}\label{7}
  \sigma_d=4\pi \frac{\omega_l}{c}\varepsilon^{1/2}_m Im[\alpha_d(\omega_l)].  
\end{equation}

Hence, the power absorbed by the bispere is  , where   is the intensity of laser radiation. Besides, the particles get polarized in the laser field and the light-induced dipoles interact with each other.
 Hence, the power absorbed by the bispere is
$ W_d=\sigma_d I$, where $I$  is the intensity of laser radiation.
 Besides that in the field of laser radiation particles are polarized
 and light-induced dipoles interact with  each other.
The above mentioned parameters are sufficient for calculation of the real part of the bisphere polarisability.

 In the case when the angle between the field intensity vector ${\bf E}$ and the bisphere axis ($\theta$)
 is  $0$ or $\pi$   (see Fig.~\ref{fig4}), the  interaction energy and the force  $F_{em}$
between particles are defined by the  expression \cite{Drach,Claro1,Claro2}:
 \begin{equation}\label{8}
  U_{em}= -\frac{2\pi}{c}I\cdot\ Re [(\alpha_d(\omega,h) - 2
  \alpha_0(\omega)], \\\  F_{em}= -\frac{d U_{em}}{dh},
\end{equation}\label{9}
 where  $\alpha_0(\omega)$ is the polarisability of an isolated particle (the second term in the expression
  for   $ U_{em}$, does not dependent on $h$, and it describes the energy of individual particles polarized in the field). Note, that under such field orientation  (at $\theta=0$) the  moment of force acting on a bisphere
  is  zero.

Thus, the accuracy of our calculations and that of the calculations in \cite{Drach} is defined by the adequacy of the Claro model
\cite{Claro1,Claro2,Claro3}.

Dispersion of the surrounding medium in the spectral range of interest (at $\lambda$=540 and 1080 nm) was taken into account when calculating polarisability of the $Ag$ bisphere (for water the difference in dielectric permeability at these wavelengths is negligible and does not fall outside the limits $10^{-2}$).

Thus, the action of laser radiation is actually reduced to two important factors:
the heating of particles  and the  appearance  of the interaction force of induced dipoles (in general case multipoles)
${F}_{em}$, that directly influences
the movement of particles in a bisphere. There are also other approaches to the optical binding
 forces calculation operating between nanoparticles in optical fields (see, e.g. \cite{Hallok,Vesperinas}).

 \subsubsection{\label{part3_1_2}Estimation of a radiation pressure}

Estimate also the role of the radiation pressure force  ($F_{pr}$).
It is important to note, that this force acts orthogonally to the bispher axes,
therefore it doesn't change a relative nanoparticles position.
For the same reason it does not have effect on the bisphere optical properties.

We estimate $F_{pr}$ using the expression:
\begin{equation}\label{aaaa8}
F_p=P_r S,\\
\end{equation}
where  $S=\pi r_0^2$  is the radiated particle surface, $P_r=2I/c$ is the radiation pressure at full  reflection of the radiation.
 Acceleration gained by a bisphere  is   $a=F_p/2m=4\pi r_0^2 I/[2c (4/3)\pi
r_0^3\rho]=3I/(2cr_0\rho)$, $\rho$  is the particle density. Taking  $I$=10$^8$ W/cm$^2$, $r_0$=10$^{-8}$ m, $\rho \simeq$ 10$^4$ kg/m$^3$
(for $Ag$) we obtain   $a= 5\times 10^7$ m/s$^2$.
Accordingly, for the  nanosecond pulse $\tau=10^{-8}$ s  at the maximum experimental
 intensity this will lead to a shift of the bisphere by $l=a\tau^2/2=2.5\times10^{-9}$m=2.5 nm. For
 $\tau=3\times 10^{-11}$ s  we have $l=2\times10^{-14}$\ m.
 If the bisphere is the part of an aggregate, the shifts will be much less.

 \subsubsection{\label{part3_1_3}Elastic interaction}

\begin{figure}
\includegraphics{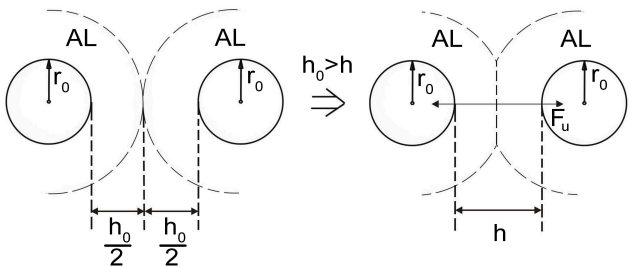}
\caption{\label{fig5}Deformation of identical particles with radius $r_0$  when their adlayers are in a contact with each other (AL is an adsorption layer).}
\end{figure}

To prevent fast aggregation of particles in hydrosols, a water-soluble polymer is often added to the initial solution before a disperse phase starts developing. In this case, as has been mentioned in the previous section, each new particle appears to be surrounded by an adsorption layer (a polymeric shell), having the thickness about $h_0/2=1\div 2$ nm and more (Fig.~\ref{fig5}). This shell reduces the Van-der-Waals attraction when particles collide, preventing coagulation (sticking) of the particles. The gap between adjacent particles in the aggregate is about $1\div5 \ $ nm  for radius $\sim10$ nm.
At such distances, the Van-der-Vaals attraction between particles ($F_v$) is already high enough for the particles to form aggregates. The key role in compensation of this attraction is played by the elasticity force ($F_e$) arising from deformation of polymeric shells of adjacent particles. To describe this force, we will use the results obtained for the problem of deformation of two balls (Fig.~\ref{fig5}), which is also known as the Hertz deformation/tangent  problem \cite{Landau_Upr}.

\begin{figure}
\includegraphics{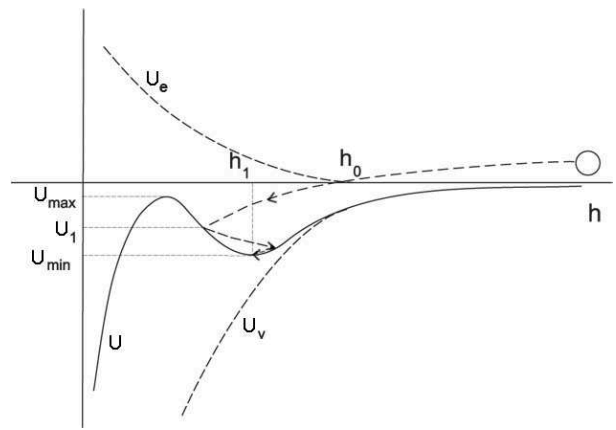}
\caption{\label{fig6} Diagram of particles capture in a potential well.}
\end{figure}

The potential energy of such an elastic interaction in the case of identical spheres of radius   $r_0$  can be described by the expression
 \cite{Landau_Upr}
\begin{equation}\label{12}
 U_e=\frac{4}{15}(h_0-h)^{5/2}\left( \frac{r_0+h_0/2}{2}         
 \right )^{1/2}\left(\frac{E_e}{1-\sigma
 _e^2}\right),
\end{equation}
where $E_e$  is the effective elasticity module of polymeric adlayer,   $\sigma _e$ is the
Poisson coefficient (the mean values  $\sim0.15$). Since $\sigma _e$  is very small, it will be neglected in our further calculations.  Unlike the classical Hertz problem, where deformation of two contacting spheres is considered, in our case the spheres contain a metal kernel inside, which at least at the initial stage of deformation does not affect the process of deformation of the external surface. When using alternative expressions for the energy of elastic (steric) interactions of polymeric adlayers of particles  (see, for example, \cite{Lewis}) it is not possible to avoid an arbitrary choice of adjusting parameters for specific types of polymeric molecules. A detailed comparative analysis of elastic (steric) interaction potentials of particles, obtained in terms of different models, will be reported elsewhere.
Consider now the process of coagulation of two identical particles (before the forces become balanced) under the action of the Van-der-Waals force
 $F_v$, the elasticity force  $F_e=-dU_e/dh$  and
the viscous friction force  $F_f$.
Fig.~\ref{fig6} illustrates the total interaction energy of particles  $U(h)=U_v+U_e$ \ ($U_v$
is the Van-der-Waals energy) as a function of the distance between their surfaces  $h$.

The particles approach each other driven by the force $F_v$. Before their polymeric adlayers come into contact this movement is counteracted by the viscous friction force  $F_f$  only and once they come into contact the elasticity force  $F_e$ adds in as well.  As a result of the particles approach a stable position of the particles in a bisphere with the gap  $h$  is achieved when the secondary potential well in  $U(h)=U_v+U_e$  (Fig.~\ref{fig6}) satisfies the following conditions:
\begin{equation}\label{13}
 U_{max}-U_1 > 0, \quad U_1=W_0-A_f, \quad
  U_{max}-U_{min}\geqslant \frac{3}{2}kT,
\end{equation}
where $W_0$ is an initial kinetic energy of particle ($W_0\approx
0$), $A_f$  is the work of the viscous friction force
  in the way from the capture of particle in the potential wall before a collision
  with a barrier separating the secondary potential wall from the main potential minimum.

\begin{figure}
\includegraphics{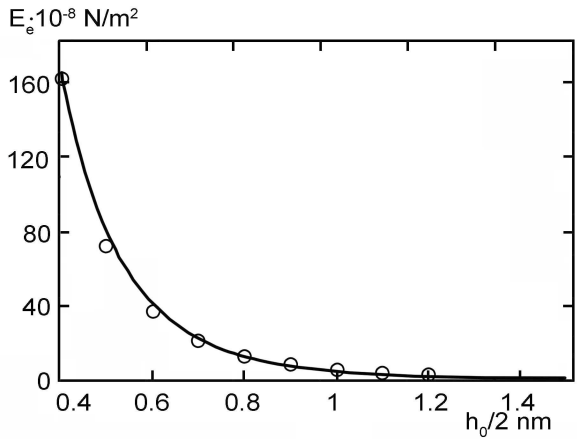}
\caption{\label{fig7}The effective elasticity module of adlayer vs the adlayer thickness
 $h_0/2$.}
\end{figure}

In Fig.~\ref{fig7} the dependence  $E_e(h_0)$, satisfying to conditions (\ref{13}) is shown.
The dependence was derived from the solution of the problem on particle movement in a potential field, taking into account the friction force and selecting appropriate values of the elasticity module. In the range
 3$\ $nm
$\geqslant h_0 \geqslant 0.8\ $nm  for the particles whith $r_0=10$nm it is well approximated by the function

\begin{equation}\label{14}
 E_e(h_0)=1.67\cdot 10 ^{8} \cdot \exp \left [ 3.09 \left (h_0/2  - 1.62
 \right )^2\right ].
 \end{equation}
Here $h_0$ is measured in nm, and  $E_e$ is done in N/m$^2$.

For the purpose of our study we will take the advantage of the data available on exactly the kind of silver sols with a casein adlayer which were used in the experiments. It should be noted, that the elasticity module values in the contact area chosen in
 (\ref{14}) for the material of the particles are 4-5 orders of magnitude higher compared to those for a free casein gel in Ref. \cite{Izmailova}.
This is explained by the fact that it is impossible, using in the calculations the reported values of  $E_e $, to account for the existence of nanoparticles aggregates with the distances between their surfaces  $h \simeq 0.5 \div 2$ nm for which the Van-der-Waals forces would counterbalance the elasticity force.

Large values of the adlayer elasticity module can be explained by the following factors. Adlayers of contacting particles in aggregates already are in a deformed state near the contact surface. Besides, the adlayer is attracted to the particle surface due to electrostatic, Van-der-Waals, ion-dipole and other types of interactions. This all means that the adlayer is in a compressed state, which increases its density in comparison with a free gel. Moreover, the gel structure changes as well due to interaction with the surface because the energy of binding of the particle surface with polymeric molecules turns out to be larger than the binding energy between the molecules (the latter accounts for the elasticity of a free gel). The thinner the polymeric adlayer, the more it is compressed and the greater is the contribution of interaction with the surface, which should result in an increased module of elasticity.
The bonds in a polymeric gel grid are destroyed during heating of particles and adlayers induced by the applied laser field, which decreases the elasticity module of the gel. This, in turn, alters the conditions of the balance of forces for bisphere particles since the elasticity force decreases. The latter leads to degradation of the secondary potential well and the particles start approaching to come into almost a full contact, i.e. to get into the main potential minimum. Thus the temperature dependence of the elasticity module is an important consideration determining the accuracy of our model. For example, the elasticity module  $E_e$ changes  from $1.3\cdot 10^4$ N/m$^2$   to
$3\cdot 10^3$ N/m$^2$  when the temperature is raised by  $30 K$
\cite{Izmailova}. In our case, however, the experiments did not reveal this kind of a strong dependence: the structure of silver aggregates did not show any noticeable changes when heated up to 370 K, which is supported by the data on the absorption spectra. This means that the elastic characteristics of polymeric adlayers of particles in aggregates are subject to a very little change, which can be attributed to their strong initial compression and higher binding energy with the particle surface.

The rate of heating is another important factor affecting the elasticity module of a polymer in the process of heating. As is known from Ref. \cite{Slutsker}, destruction of a polymeric grid occurs within terminal time  $\tau _d$. So, if it does not take place within the time equal to the laser pulse duration there will be no changes in the elasticity module. According to the theory \cite{Frenkel},  an average waiting time for local energy fluctuation $E_f$ (in a single atom or a small group of atoms bound into a grid point) depending on the temperature is found as
\begin{equation}
\tau _f\sim\tau_0\cdot \exp({E_f/kT}),
\end{equation}
 where $\tau_0\approx 10^{-12}\div10^{-13} $s. The bond with the potential barrier $U_f$  will be destroyed when $E_f\geqslant U_f$.

The average waiting time for this to happen is
\begin{equation}
\tau _b\sim\tau_0\cdot \exp({U_{f}/kT}),
\end{equation}
which is the time required for a polymer to relax into a new stable state (with a smaller number of bonds between the molecules)  corresponding to the changed temperature.

Proceeding from the fact that fractal aggregates remain stable when heated up to
  $100 ^o C$ ($\sim 370 \ $ K), we estimate  $U_f$.
The values of   $U_f$ obtained for various relaxation times at  $\sim 370 \ $
K and $\tau _0 =10^{-12} \ $ s are summarized in the Table 1.

\begin{table}
\caption{\label{tab:table1}The potential barrier for different relaxation times  }
\begin{ruledtabular}
\begin{tabular}{cccc}
$\tau_b, $ s  &$60$ &600 &3600\\
\hline
$U_f$, eV & 1,04 & 1,12 & 1,17\\
\end{tabular}
\end{ruledtabular}
\end{table}

Assuming the average value $U_f=1.1 \ eV$ (which, incidentally, is in a good agreement with a similar value for polymeric media), we find the critical temperatures ($T_{cr}$) below which the times $\tau
_b =10^{-8}$ and  $10^{-11} \ $s are not long enough for relaxation to occur. We obtain  $\tau _b =10^{-8} $ s --- $T_{cr}=1385 \ $K; \ $\tau
_b =10^{-11} \ $s ---
 $T_{cr}=4070 \ $ K. Theoretically, once these temperatures are reached, the polymer elasticity module should drop dramatically. In our opinion, the change of $E_e$   with temperature can be assessed with more accuracy in terms of the fluctuation frequency $\nu_r$:
\begin{equation}
  \nu_r={\tau_b^{-1}}={\tau_0^{-1}}\exp({-U_f/kT}).
\end{equation}

In this case it is reasonable to draw an analogy with the chemical reaction at the rate $\nu_r$.
Then  the change of $E_e$ can be expressed as
\begin{equation}
  \frac {dE_e}{dt}=-\nu_rE_e.
\end{equation}

It is easy to check, that for a short time  $\Delta t<\nu_r^{-1}$ a  relative change satisfies the equation:
\begin{equation}
  \frac {\Delta E_e}{E_e}\approx  \frac {\Delta t}{\tau_0}\exp({-U_f/kT}).
\end{equation}

 If  ${\Delta t}\cdot {\tau_0^{-1}}\cdot \exp[{-U_f/kT}] \ll 1$,
 then  $ \Delta E_e \ll E_e$, i.e. there is virtually no change.
For this purpose
 the condition $T< U_f/{\ln
(\Delta t / \tau_0)}$  should be satisfied which corresponds to the definition   $T
_{cr}$.

Note that the properties of a polymer can be strongly modified by other factors as well. For example, at the particle surface temperature
 $T \gtrsim 600$ K, explosive vaporization takes place in the adjacent layer of overheated water with a resultant dehydration of the polymeric layer   \cite{Pustovalov}.

It is therefore necessary to take into consideration the whole variety of factors affecting the elastic interaction of particles in a bisphere. Besides, the repulsive interactions between particles are also caused by the interaction of electric double layers (EDL) consisting of electrolite ions which are adsorbed by particles along with polymeric molecules.

 \subsubsection{\label{part3_1_4}Electrostatic interaction}

\begin{figure}
\includegraphics{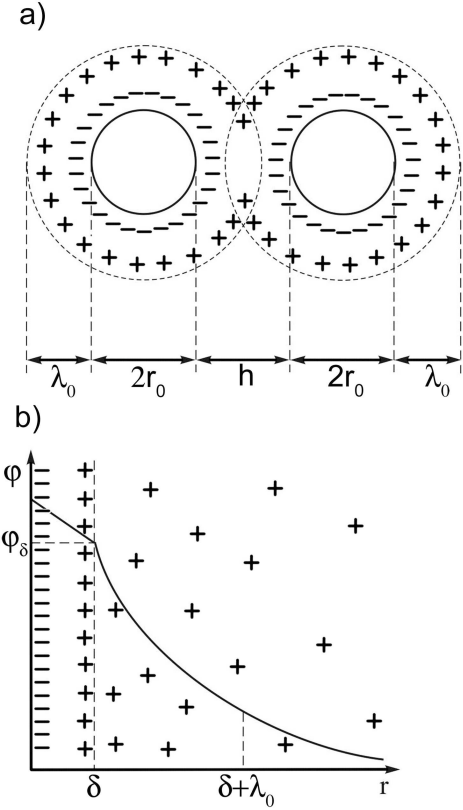}
\caption{\label{fig8}Electrostatic interaction of two spherical nanoparticles:
   a) overlapping of double electric layers (EDL) when particles collide; b) the structure of EDL. }
\end{figure}

Electrostatic interactions of spherical nanoparticles in a bisphere are generally described in terms of the Derjaguin-Landau-Verwey-Overbeek (DLVO) theory (see for example
 \cite{Frolov}). According to this theory, each particle
in electrolite solution is surrounded by an electric double layer consisting of electrolyte ions. Electrostatic interaction is virtually absent when the distance between the surfaces of two particles exceeds the total thickness of these two layers  ($2\lambda _0$), and at shorter distances these layers overlap (Fig.~\ref{fig8}a).  Due to a partial displacement of electric charges from the external part of EDL (a counterion layer), there is an electrostatic repulsion of the internal, dense parts of the EDL particles (layers of potential-determining ions). Since concentration of electrolite ions
can be quite high
(up to  $10^{-3}\div \cdot10^{-1}$ mol/l)
 in polymer-containing silver hydrosols, the particle surface adsorbs polymers as well as ions which form EDL. The thickness of the dense part of EDL is defined by the ions concentration and depending on this concentration it can either exceed the thickness of the polymeric part of the adlayer or be localized inside that part. The latter corresponds to our experimental conditions ($\lambda _0\leqslant 1$\ nm));
 electrostatic repulsion takes place only in case of appreciable deformation of polymeric shells when the interparticle gap reaches the total thickness of the dense parts of EDL of neighbouring particles.

The potential energy of Coulomb interaction $U_q(h)$  of overlapping EDLs of  spherical particles of radius $r_0$
  is defined as  (see, for example, \cite{Frolov,Zontag,Ansell,Lewis,Enustun}):

\begin{equation}\label{21}
U_q(h)=2\pi\varepsilon_0\varepsilon
r_0\varphi^2_\delta\ln\left[1+\exp({-h/\lambda_0}]\right),
\end{equation}
where $\varepsilon_0$ is dielectric constant,  \ $\varepsilon$ is a  relative dielectric permittivity of the surrounding medium,
$\lambda_0$ is the  Debye-H\"{u}ckel radius, $\varphi_{\delta}$ is the potential at the border of the Helmholtz layer
(the layer thickness  doesn't exceed the size of hydrated counterions constituting the layer).
The latter is  the parameter describing the EDL on a particle surface.

Unlike the Youkawa potential (see, for example, \cite{Sauer}), Expression  (\ref{21}) is more convenient for estimation of $U_q(h)$ as it contains the parameter $\varphi_{\delta}$. The potential $\varphi_{\delta}$  for a low electrolite concentration  is close to be equal to electrokinetic  ($\zeta$) potential
 defined experimentally by  means of  electrophoresis \cite{Frolov,Enustun} (the potential at the glide plane of a nanoparticle moving in an applied electric field).
A further advantage of using this expression is that it does not require data on the effective screened charge of particles in the layer of potential-determining ions.

For estimation of the EDL parameters, described in the  Stern theory (see for example \cite{Frolov}),
consider its structure. In Fig.~\ref{fig8}b characteristic EDL borders with a potential of the Helmholtz layer  $\varphi_{\delta}$ and  $\lambda_0$   (the characteristic extension of the diffuse part of a counterions layer)  are shown (at the  distance
where the potential of the layer of potential-determining ions located on a particle surface  falls by $e$ times).
In order to estimate the EDL parameters described in the terms of the Stern theory (see for example \cite{Frolov})), consider the structure of EDL (Fig.~\ref{fig8}b). Typical EDL boundaries with the Helmholtz layer potential $\varphi_{\delta}$  and the typical size of the diffuse part of a counterions layer $\lambda_0$   are shown in Fig.~\ref{fig8}b.

 In the case of spherical particle $\varphi_{\delta}$  can be found as \cite{Frolov}:
 \begin{equation}\label{22}
\varphi_\delta= -\frac {q_0}{4\pi \varepsilon _0 \varepsilon  r_0
}+ \frac {q_{c}}{4\pi \varepsilon _0 \varepsilon ( r_0 +\delta )},
\end{equation}
 where $q_0$  is  the total charge of potential-determining ions adsorbed on the particle surface,
$q_{c}$ is the charge of  counterions  in the Helmholtz layer.

According to the Stern theory, it's determined by \cite{Frolov}:
\begin{equation}\label{23}
{q_c  = \frac{{q_\infty  }}{{1 + x^{ - 1} \exp ( A_t /T)}},
 \ A_t= \frac{F \cdot z \cdot \varphi_{\delta}  + \Phi}{R}.}
\end{equation}

Here $q_\infty$ is the  charge limit for counterions in the Helmholtz layer   $q_\infty\leqslant q_0$, \ $x$  is the dimensionless parameter
equals to the mole fraction of counterions in the solution, \ $F$ is the Faraday constant, \ $z$  is the counterion charge ($z$=1),
  \ $\Phi$ is the energy of specific adsorption, $R$ is the absolute gas constant.

Taking into account that $\delta \ll r_0$ ($\delta$ is of the order of the counterions sizes) and assumung $q_\infty \approx q_0$,
  we obtain  from (\ref{22}) and (\ref{23}):

\begin{equation}\label{24}
\varphi_{\delta}= -\frac{q_0}{4\pi \varepsilon _0 \varepsilon r_0}
\cdot \frac{\exp{[(F\cdot z\cdot
\varphi_{\delta}+\Phi)/RT]}}{x+\exp{(F\cdot z\cdot
\varphi_{\delta}+\Phi )/RT]}}.
\end{equation}

   Values   $q_0$ and $\Phi$  can be derived from the results of \cite{bookKarpov,CJ2002}, where for the same type of hydrosol
at the given electrolite ($NaOH$)  concentration  ($c_m$)   and  $T=300 $ K the following values were obtained:
$\zeta=-0.064 \ V$  at $c_m\backsimeq 10^{-3} \ $ mol/l    and $\zeta=-0.034 \ $V
at  $c_m\backsimeq 10^{-2} \ $ mol/l. Bearing in mind that  $\zeta$-potential at low values of $c_m$ is
equal to $\varphi_{\delta}$ \cite{Frolov,Enustun} we can use these values  to estimate $q_0$ and $\Phi$.
Using (\ref{24}), we find  $\Phi=-2.02\cdot 10^4 \ $ J/mol, \
$q_0=9.8\cdot10^{-18}\ $ C, \ which agrees with other estimates of the charge of a particle \cite{bookKarpov,CJ2002}.

    Moreover, an estimate  for  $\lambda_0$ at $c_m\backsimeq 10^{-1} \ $ mol/l and $T=300 $ K is available  \cite{bookKarpov,CJ2002}, which is  $\lambda_0
\approx 10^{-9} \ $ m.  Since  $\lambda_0 \propto \sqrt{T}$, we obtain the expression for $U_q(h)$
with quite specific  values $\varphi_\delta (T)$ and $\lambda_0(T)$ at $c_m=10^{-1} \ $ mol/l

\begin{eqnarray}\label{25}
\begin{array}{l}
 \varphi _\delta   \approx  - 0.11[1 + 6\exp ( - 1.16 \cdot 10^4 \varphi _\delta  /T)]^{ - 1} (V), \\
 \lambda _0  \approx 1.7510 \cdot 10^{ - 2}\sqrt{T/c_m}  (nm). \\
 \end{array}
\end{eqnarray}

 In particular, at $T=300 $ K we obtain
 $\varphi_{\delta}\approx-0.0108 $ V  for
 $c_m=0.1 $ mol/l,
 corresponding to the experimental conditions in the used  silver hydrosols.

Expression (\ref{25}) is valid for a steady  EDL.
In the  process of photomodification  the particles and
surrounding medium get  heated under  the action of laser radiation. The EDL parameters change with increasing the temperature.
A  question arises whether  EDL will have enough time to achieve a steady state during a heating?

Let us estimate the  time of achieving a balance of charges in the layer of $\lambda_0$ thickness:
$\tau({\lambda_0}) \sim {\lambda_0}/{\rm v _i}$, $\rm v _i$ is an average  velocity of counterions in the changing
electrostatic field of a particle ( $q_0$ is assumed to be independent of $T$). In turn
\begin{equation}
 \rm v_i\sim B_i |E|, \ E \leqslant E_{\delta} =
-\frac{(q_0-q_{c})}{4\pi \varepsilon_0 \varepsilon {r_0}^2},
\end{equation}
where  $B_i$ is the counterions mobility in the medium, $E$ is an intensity of electric field generated
by the Helmholtz layer of  ions in the diffuse part of EDL,   $E_{\delta}$  is the electric field intensity
at the external boundary of EDL. 

Taking into account (\ref{24}), we have an estimate for $E_{\delta}$ è $\tau_{\lambda}$
\begin{equation}
|E_{\delta}|\sim \frac{\varphi_{\delta}}{r_0}, \                           
\tau_{\lambda}\gtrsim \frac{\lambda_0 r_0}{B_i|\varphi_{\delta}|}.
\end{equation}

    At $r_0=10^{-8}$ m, $\varphi_{\delta}\sim 0.01 \ $ V and
$\lambda_0\simeq10^{-9}\ $ m  (for $c_m=0.1\ $ mol/l) we obtain
$|E_{\delta}|\simeq10^6 \ $ V è $\tau_{\lambda}\gtrsim2\cdot10^8\ $
s  for a mobility of counterions in   water $B_i\sim4\cdot10^{-8} $
m/(V$\cdot $s). At  the same time laser pulses used in experiments have durations
$\sim10^{-8}$ and \ $10^{-11} $ s.

Thus, the structure of EDL changes very little under applied laser field when the temperature of the surrounding medium changes. Therefore in our further discussion the properties of EDL particles will be assumed to remain unaffected by laser radiation. Moreover, we should also bear in mind that the presence of a polymer in the adlayer can severely decrease the mobility of counterions. Hence the time required for EDL restructuring will increase.

\vspace{1cm}
 \subsubsection{\label{part3_1_5}The Van-der-Waals interaction}

Dispersion forces of the Van-der-Waals attraction are one of the key factors determining movement of particles in a bisphere during laser irradiation. We will employ the Hamaker and de Boer theory (see, for example, \cite{Zontag,Ansell}) to describe these forces. According to the theory, the interaction energy of two spherical particles with the radius  $r_0$  is described by the following expression:
\begin{equation}
 U_v = -\frac{A_{H}}{6}\left [ \frac {2r_0^2}{r^2_{12}-4r^2_0}
 +  \frac {2r_0^2}{r^2_{12}}+ \ln \left ( 1- \frac{4r_0^2}{r^2_{12}}\right )
 \right],
\end{equation}
where $r_{12}=2r_0+h$  is the distance between the centers of  particles, $A_{H}\approx 50  kT_0$ is
the  Hamaker constant for metal hydrosol \cite{Derjagin}, In the calculations, a limitation is made on the value
 here  $T_0=300 $ K, $k$ is the  Boltzman constant.
In calculations the limitation is made on the value  $({r^2_{12}-4r^2_0})/4r^2_0>0.0001$.
Thus, the Van-der-Waals force acting on each particles is equal to  $F_v=-
dU_v/dh$.

\subsubsection{\label{part3_1_6}The force of viscous friction}
When particles move in a bisphere they experience counteraction from the medium (in hydrosols - from water) - the force of viscous friction. This force is estimated by the Stocks formula:

\begin{equation}
 F_f=6\pi \eta r_0V_p,
\end{equation}
where  $\eta $ is the  viscosity of a  medium, $V_p$  is  velocity of  particles in a bisphere relative to its  center of  mass.

\subsubsection{\label{part3_1_7}Heating of particles with consideration of
 heat emission into surrounding medium}

The primary effect of absorption of laser radiation by metal particles of a biosphere is an increased internal energy of the electron gas.  Energy transfer from electrons to the metal crystal lattice occurs rather slowly because of the big difference in the mass of electrons and ions. Equilibrium times in the electronic  $\tau _e$  and ionic   $\tau _i$  subsystems are much shorter than the equilibrium time between electrons and the lattice. So it is possible to introduce two subsystems: electronic and ionic. For  $\tau _e$  and $\tau _i$  we obtain the following estimates:
\begin{equation}\label{31}
  \tau _e \sim (\sigma _q \cdot \rm v_e \cdot N)^{-1}; \quad
  \tau _i \sim (\sigma _q \cdot v_i \cdot N)^{-1},
\end{equation}
where $\sigma_q$   is  cross-section of Coulomb collisions, $\rm v_e$, \
$\rm v_i$ are the average electron and ion velocities, respectively, $N$ is the  concentration of particles.

With the typical values of  of electron (ion) concentration in a particle being
$N \sim 10^{28}$ m$^{-3}$, for $kT\simeq 1$~eV   we have $\sigma _q \sim 10^{-17}$ m$^2$, \ $\rm v_e = 6\cdot 10^5$ m/s, \
 $\rm v_i \sim \rm v_e\sqrt {m_e/m_i}$. Using (\ref{31}), we find
  $$
 \tau _e \sim  10^{-17} s; \quad  \tau _i \sim  10^{-14}\div  10^{-15} s.
 $$

     That is equilibrium times are much shorter in comparison with pulse duration and typical
     heating times  $ \tau _h \geqslant10^{-12} $ s  (under our conditions), so it is possible to introduce temperatures   and  $T_e$ for electronic and ionic $T_i$ subsystems, respectively. Under these conditions we can write down the following equations for $T_e$  and $T_i$  \cite{Anisimov}

 \begin{eqnarray}
C_e \frac{dT_e}{dt}  =  -g(T_e-T_i) + \frac{\sigma_d I}{v_d}, \nonumber  
\\[0.5cm]
C_i \frac{dT_i}{dt}  =  g(T_e-T_i) -q_l,
\end{eqnarray}
where $C_e$, $C_i$  are electronic and lattice thermal capacities, respectively,
\ $g$ is the rate of energy exchange between subsystems, $v_d$  is the  bisphere volume.
These values are taken from \cite{wright}, where

$ C_e=68\cdot T_e$ J$\cdot$m$^{-3}\cdot $K$^{-1}$; \quad $C_i=2.5\cdot10^6 $ J$\cdot$m$^{-3}\cdot $K$^{-1}$ 
; \quad  $g
=4\cdot10^{16}$ J$\cdot$m$^{-3}\cdot $K$^{-1} \cdot $s$^{-1}$. 

 The term $\sigma I/v_d$  account for  energy gain by the electronic subsystem (to unit  volume)
 due to  absorption of laser radiation with intensity  $I$, \ $q_l$ is heat loss from a particle into the environment.

Definition of $q_l$ is rather difficult problem for which  a non-stationary problem
of heat conductivity has  to be solved first with consideration for the phase transition "liquid-vapour"  \cite{Pustovalov,Belousova}. This essentially complicates understanding of the particles kinetics in a bisphere under laser radiation heating. Therefore we will use the following simple model. The heat loss per unit volume is described as
\begin{equation}\label{q_l}
  q_l = \frac{j \cdot S}{v_d} = \frac{\mbox{\ae} \cdot \rm{grad} (T_i) \cdot S}{v_d},
\end{equation}
where   $S$ is the surface of a particle, $j$ is the heat flux density, \
$\mbox{\ae}$ is the heat  conductivity factor of water.

\begin{figure}
\includegraphics{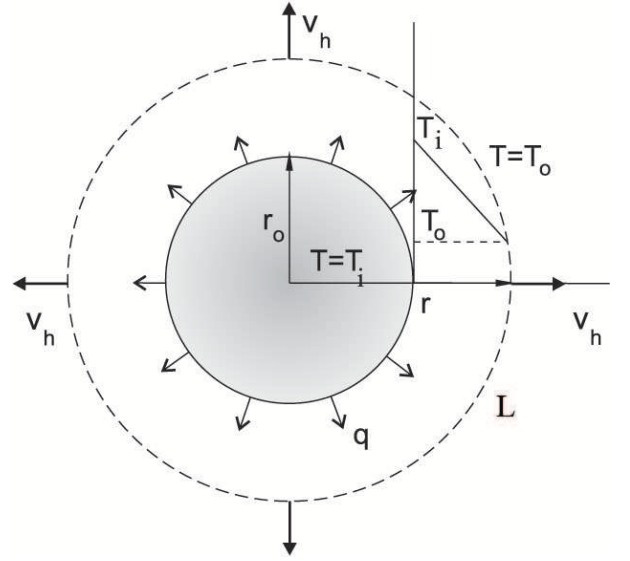}
\caption{\label{fig9}Heat exchange between the particle and the surrounding medium}
\end{figure}

Let $L$ is the boundary of heated area in surrounding liquid (a water) (Fig.~\ref{fig9}).

The  speed of movement of this border $\rm V _h \sim
\sqrt{{\emph{a}_m}/{t}}$, \ $a_m= {\mbox{\ae}_m}/c_{mo} \rho _m$ is a thermal diffusivity coefficient of water,
 \ ${c_{mo} , \ \rho
_m}$  is a specific thermal capacity and density, respectively.

Then the border $r$ of a heated area is defined by the equation
\begin{equation}
  \frac{dr}{dt}=\sqrt{\frac{a_m}{t}}, \qquad
r= r_0 + \frac{1}{2}\sqrt{a_m \cdot t}.      
\end{equation}

Now we can  estimate $\rm{grad} (T)$:
\begin{equation}
 \rm{grad} (T) \approx  \frac{T_i - T_0}{r-r_0}=2\sqrt{\frac{
\ c_{mo}\ \rho_m}{\mbox{\ae}_m t}} (T_i-T_0),            
\end{equation}
where $T_0$ is the temperature of water near the border of heated area ($T_0=300 \ $K).

Using (\ref{q_l}) we find  $q_l$:
\begin{equation}
 q_l= \frac{6}{r_0} (\mbox{\ae}_m \ c_{mo} \ \rho_m)^{1/2}
 t^{-1/2}(T_i-T_0).            
\end{equation}

 In the model under discussion  the parameters of a medium are assumed to be invariable, although they depend on the medium temperature,
 and also change during the  phase transition "liquid-steam". However, as follows from \cite{Pustovalov},
 phase transition can occur at  $T\approx600 \ $K. That means that the liquid although overheated maintains its aggregative state.

\subsubsection{\label{part3_1_8}Kinetic  equations for a  bisphere in the  process of photomodification}

\begin{figure}
\includegraphics{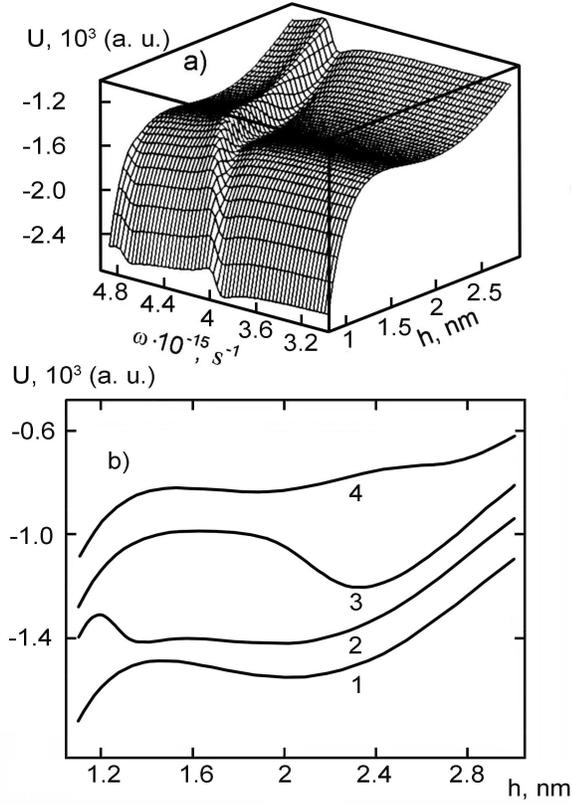}
\caption{\label{fig10}Potential surface $U(h,\omega)$  (a) for a bisphere with the initial distance  $h_1=2\ $nm.
Potential curves (b) for different $\omega$:\ 1
    ---
    $2.5 \cdot 10^{15}$ \ s$^{-1}$; 2 ---  $4.2 \cdot 10^{15}$\ s$^{-1}$; 3
    ---  $4.45 \cdot 10^{15}$\ s$^{-1}$; 4 ---  $4.53 \cdot 10^{15}$\
    s$^{-1}$. For better illustration Curves 1-4 are shifted along the vertical axis relative each other.}
\end{figure}

As follows from the above discussion, the change of the bisphere characteristics  under the action of optical radiation,
which we will call "photomodification", depends on many factors such as the laser pulse energy and duration, the radiation frequency and also the initial state of the bisphere (the interparticle distance). The shape of the potential energy surface shown in Fig.~\ref{fig10} (for the laser radiation intensity  $I=10^7$ W/cm$^2$    and the initial distance between particle surfaces   $h_1= 2$ nm) can give an idea of the movement of particles in a bisphere.

The potential surface in absence of  radiation  is defined by the Van-der-Waals, elastic and electrostatic interactions.
In Fig.~\ref{fig10}a the surface section with  $\omega \leqslant 3.6 \cdot
10^{15}$ s$^{-1}$  describes a surface in absence of radiation
(Curve 1 in Fig.~\ref{fig10}b) where the potential wall is observed. The fold in the field $\omega>3.8\cdot 10^{15}$s$^{-1} $ is the result of the light-induced particle interactions. These interactions, as is shown in Fig.~\ref{fig10}b, can lead to a shift of the potential well (Curves 2,3) or to its disappearance. That is, the particles which were in the stable position $h_1$  start moving to a new stable position under the action of radiation. Their resonant frequency will be shifted as well both to the low-frequency range (Curves 2, 4 in Fig.~\ref{fig10}b) and to the high-frequency range (Curve 3).

The situation, in actual fact, is more complicated. The particles and the polymeric layer are heated up by laser radiation to result in the change of
the module of elasticity and of the shape of the potential surface. The potential surface shown in Fig.~\ref{fig10} corresponds to the beginning of influence of laser radiation and it can describe the movement of particles at the initial stage.
In order to describe the bisphere behaviour (the interparticle distance, resonant frequencies, temperature), it is necessary to solve the system of kinetics equations in which the factors analyzed in the previous sections are taken into account. This system can be written down as follows:

\begin{eqnarray}\label{38}
&\frac{{dh}}{{dt}}= 2V_p , \quad
 \rho V\frac{{dV_P }}{{dt}} = F_p  - F_f& \nonumber \\
  \nonumber\\
 &\frac{{dE_u }}{{dt}} = - \nu _r E_u&  \nonumber \\
  \nonumber\\
 &C_e \frac{{dT_e }}{{dt}} =  - g(T_e  - T_i ) + \frac{{W_d }}{{2V}}&\nonumber \\
  \nonumber\\
 &\frac{{dQ_i }}{{dt}} = gV(T_e  - T) + q_l V,& \nonumber \\
  \nonumber\\
 &F_p  =  - \frac{{\partial U}}{{\partial h}},&\nonumber\\
  \nonumber\\
 &U = U_\upsilon   + U_{em}  + U_e  + U_q ,&\nonumber \\
  \nonumber\\
&T_i  = \frac{{Q_i }}{{C_i V}}H(Q_1  - Q_i ) + &\nonumber\\
 \nonumber\\
&+ \frac{{Q_i  + Q_2 }}{{C_i V}}H(Q_i  - Q_2 ) + T_L H(Q_i  - Q_1 ), &\nonumber\\
  \nonumber\\
 &Q_1  = \frac{1}{2}C_i VT_L, \quad Q_2  = Q_1  + L_m \cdot V. &\nonumber\\
\end{eqnarray}

Here $(1/2)V$  is a particle volume, $Q_i$  is a thermal energy of a particle,
$T_L$  is the melting temperature of silver, $L_m$  is the melting heat per $Ag$ unit volume, $H(x)$ is the Heaviside   function.

 Here we assume, that the thermal capacity $C_i$ does not depend on  temperature.
\section{\label{part4}Results}

In this section only typical examples of solution of Equation system (\ref{38}) will be given. A detailed analysis of these solutions for various individual parameters of the system, and for their arbitrary combinations will be presented elsewhere.

\begin{figure}
\includegraphics{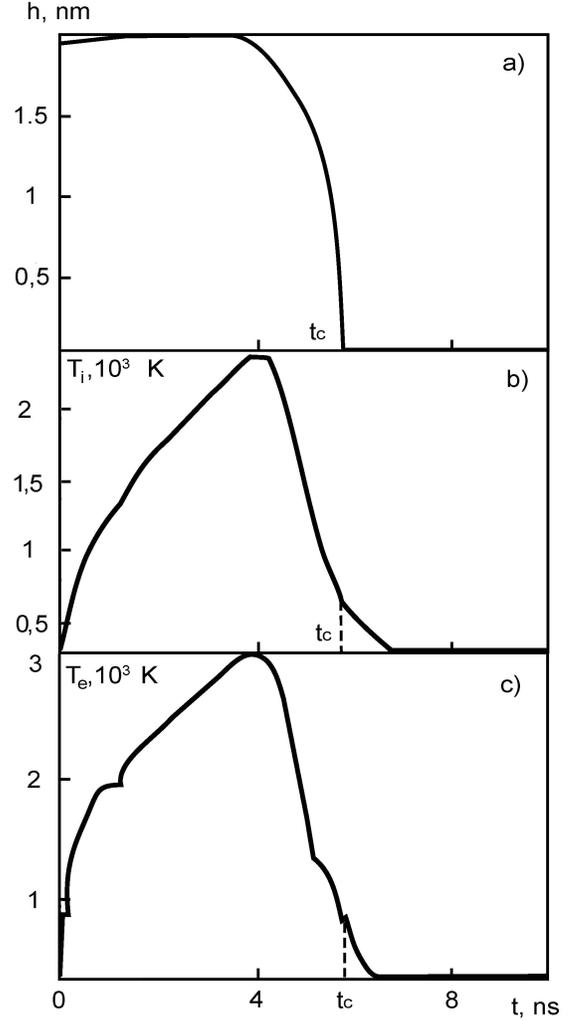}
\caption{\label{fig11}The kinetics of $h, T_i, T_e$ under applied  nanosecond laser pulse   ($I=10^6$ W/cm$^2$, $\tau=10^{-8}$ s)
in a bisphere with initial interparticle distance   $h_1=2$ nm ($t_c$ is the moment of collapse).}
\end{figure}

\begin{figure}
\includegraphics{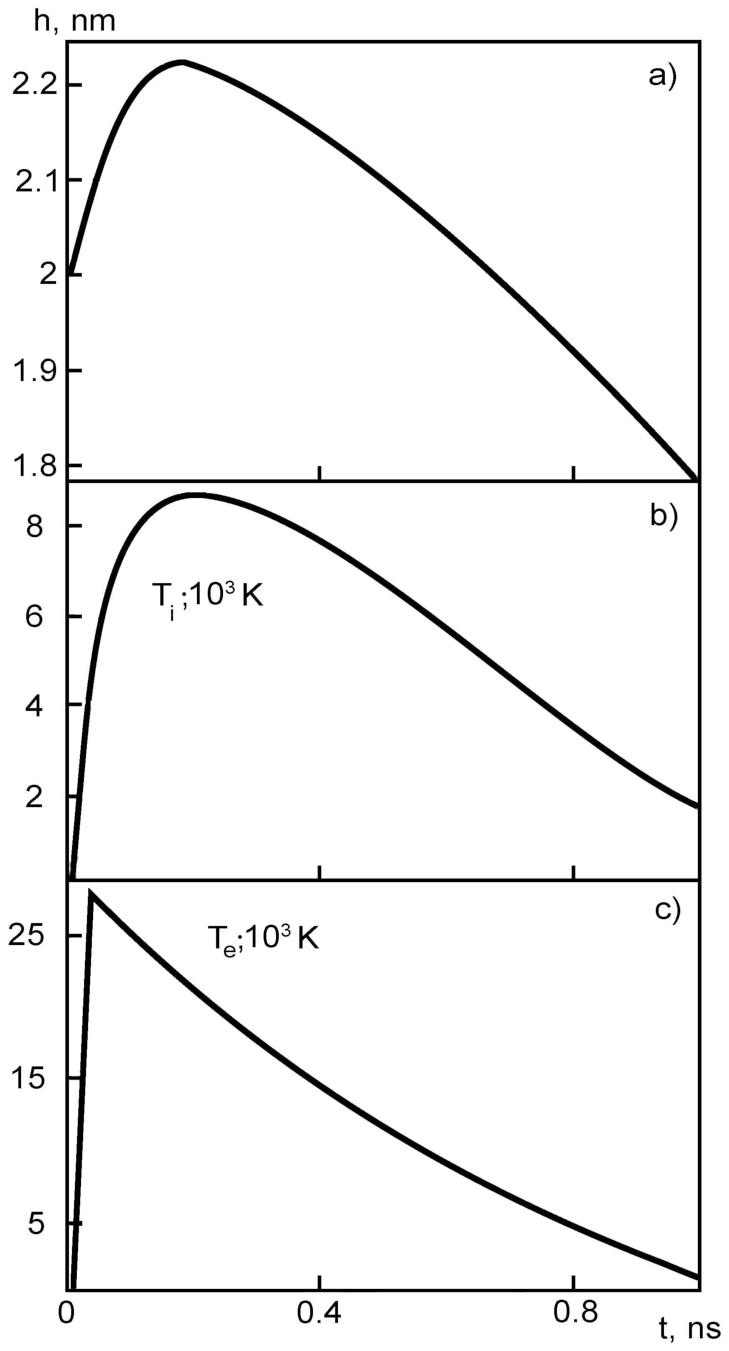}
\caption{\label{fig12}The kinetics of $h, T_i, T_e$ under applied  picosecond laser pulse   ($I=3.3 \cdot 10^8$ W/cm$^2$ , $\tau=3 \cdot 10^{-11}$ s)
in a bisphere with initial interparticle distance   $h_1=2$ nm.}
\end{figure}

The equations (\ref{38}) were solved numerically for various initial interparticle distances in bisphere,
as well as for various intensities and  frequencies of laser radiation.  Fig.~\ref{fig11}b,~\ref{fig12} show the calculated behaviour  of the interparticle distance $h$,
 ionic lattice temperatures  $T_i$   and electron  $T_e$  for the initial value  $h_1=2$ nm, the laser radiation
frequency   $\omega _{l} =\omega _{01}$
($\omega _{01}$ is and the initial resonant frequency of the bisphere) and the pulse  energy  $W=10$ mJ (the rectangular pulse).
Note that the results in Fig.~\ref{fig11} were obtained  for the pulse duration $\tau _i=10$ ns  (and intensity $I=10^6$ W/cm$^2$)
and the results in Fig.~\ref{fig12} were obtained for $\tau _i=30$ ps   ($I=3.3 \cdot 10^8$ W/cm$^2$).

As can be seen from Fig.~\ref{fig11} the particles of bisphere and surrounded medium are quickly heating up, and as a result of it
 the elasticity module of a polymeric adlayer decreases that leads to the collapse of bisphere  ($h\rightarrow0$)
 under the action of the Van-der-Waals forces (to approaching  of particles to minimal interparticle distance).

The frequency $\omega_1$ ($\omega_1=\omega_{01}(t), \omega_2=\omega_{02}(t)$ for $t>$0) decreases with the interparticle distance $h$. Radiation with the frequency \ $\omega _{l}$ falls out of resonance, the radiation absorption drops and the bisphere cools down.
By the moment of collapse ($t_c$) the lattice temperature  $T_i$  becomes less than
temperatures of melting $T_i$  of the particle's material.

That is, the resonant frequencies $\omega_1 $   and  $\omega_2 $  decrease as a result of such influence. In this case, resonant domains in a large aggregate are subjected to restructuring, which shows itself as photomodification of the aggregate. As has been mentioned before, the major contribution to photomodification is made by changes in the local structure of aggregates, or, more precisely, by changes in the structure of resonant domains.
In the approximation of domains by bispheres, the key role is played by the heating of bispheres and by the Van-der-Waals forces. The effect of light-induced forces  $F_{em}=-\partial U_{em}/\partial h$ on bisphere particles at the given radiation intensities and mean distances ($h\leqslant 2$ nm) is negligible. Note also that at the given intensities the electron temperature (Fig.~\ref{fig11}c) is just above  the temperature of the lattice.

An entirely different picture is observed in the case of a picosecond pulse (Fig.~\ref{fig12}) with   $\tau _i =30$ ps.
As can be seen from Fig.~\ref{fig12}a, the interparticle distance $h$  starts growing as soon as the picosecond pulse is applied and it keeps growing after the pulse is over. This is associated with the light-induced multipole forces $F_{em}$ increasing with the radiation intensity and at  $\omega
_{l}=\omega_1$   their effect is to make the particles go apart ($\partial U/
\partial h>0$  as shown in Fig.~\ref{fig10}b, Curve 3. The particles continue moving away from each other for a while after the action of the pulse is over gradually slowing down due to friction and the Van-der-Waals force and eventually they start collapsing.

Another distinctive feature of the process is a higher temperature of the ion subsystem which is still heated by electrons after the pulse due to the considerably higher electron temperature. As can be seen from Figs.~\ref{fig12}b and \ref{fig12}ñ, unlike nanosecond pulses, the picosecond pulse duration is not long enough for temperatures of electrons and ions to become balanced. On top of that, high temperatures at picosecond pulses are explained by the fact that the $h$ distance changes very little within the pulse duration time and $\omega _{l}\approx \omega_1$, which is different from nanosecond pulses where the bisphere collapses and falls out of resonance with the laser radiation in the middle of the pulse duration.

\begin{figure}
\includegraphics{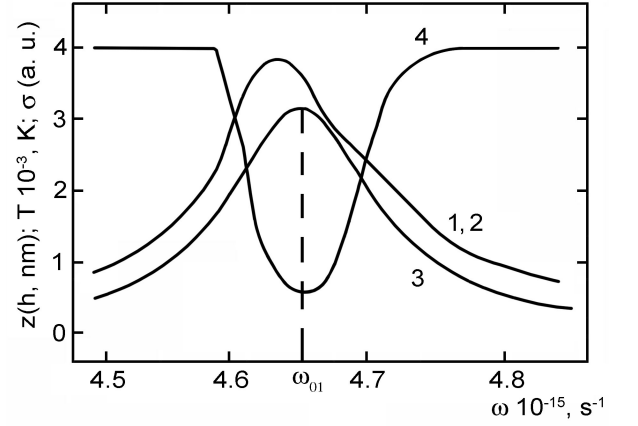}
\caption{\label{fig13}Temperatures and the distance $h$  vs laser radiation frequency    ($\omega=\omega_l$)
   for the initial  $h_1=4$ nm. 1 --- $T_{max}$, 2 ---  $T_c$, 3 --- $\sigma(\omega)$, 4 ---  $z=10(h-3.6)$  ($h$ --- in nm).}
\end{figure}

\begin{figure}
\includegraphics{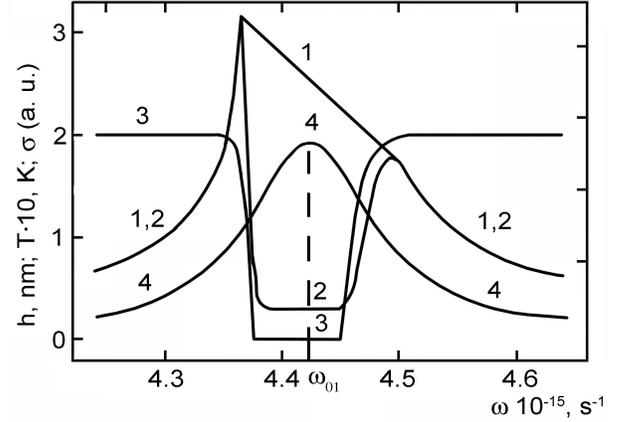}
\caption{\label{fig14}Temperatures and the distance $h$  vs laser radiation frequency   ($\omega=\omega_l$)
    at initial value  $h_1=2$ nm. 1 --- $T_{max}$, 2 ---  $T_c$, 3 --- $h$, 4 ---  $\sigma(\omega)$.}
\end{figure}

In addition we studied the dependence of  the photomodification kinetics  of bisphere on the frequency of laser radiation. The values of maximal temperature  $T_{max}$, as well as  the temperature ($T_c$) and  interparticle distance $h_c$  at the end of a pulse $\tau =10$ ns
and $I=10^6$ W/cm$^2$ for different initial interparticle distances are shown in Fig. \ref{fig13}, \ref{fig14}.
We also studied dependence of the bisphere photomodification kinetics on the laser radiation frequency. Figures \ref{fig13}, \ref{fig14} show the maximum temperature  $T_{max}$, the collapse temperature $T_c$ and the interparticle distances $h_c$  at the end of the pulse  $\tau =10$ ns  and   $I=10^6$ W/cm$^2$  calculated for different initial interparticle distances.

\begin{figure}
\includegraphics{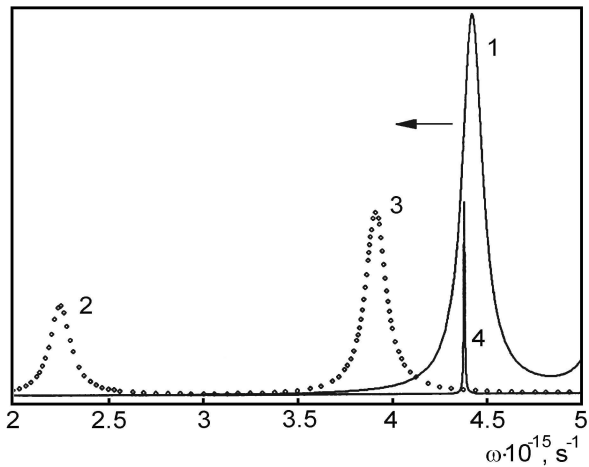}
\caption{\label{fig15}Changes in a plasmon absorption spectrum of bisphere when particles collapse: Curves correspond to:  1  --- the spectral band with resonant frequency $\omega_{01}$   before collapse,  2 ---  the spectral band with resonant frequency $\omega_1$   after collapse, Curve 3 --- short-wavelengths resonant band ($\omega_2$)  after collapse (before collapse  $\omega_2$ almost coincides with $\omega_{pl}$), Curve 4 --- the laser radiation spectrum with a frequency $\omega_l$.}
\end{figure}

As one can see from the figures, with the laser radiation frequency  $\omega _{l}$ coming close to the resonant frequency of the bisphere, the temperature of partricles is growing and the distance between the particles $h$ is reducing and the dependence $h_c(\omega_{l})$ exhibits a dip. This dip however is insignificant at large initial distances $h_1$ (Fig.~\ref{fig13}) and it is virtually symmetrical relative to $\omega _{01}$. Such an insignificant change in  $h_c$ is due to the Van-der-Waals force being relatively weak at large $h_1$. The insignificant change of $\omega_{1}$  does not lead to a shift of   therefore heating of the bisphere continues throughout the pulse duration. The temperature maximum is achieved at the end of the pulse ($T_{max}=T_c$).

Quite a different situation develops in the case of small initial distances (Fig.~\ref{fig14}). In this case the Van-der-Waals force is stronger compared to the first case and the bisphere, when heated to high enough temperatures, collapses within the pulse duration. The resonant frequency $\omega_1$  is shifted and heating of particles ceases. Thus, the maximum temperature is achieved before the collapse, after which the bisphere starts cooling down ($T_{max}>T_c$).

Another distinctive feature is the asymmetry of the temperature curves  $T_{max}(\omega_l)$  and $T_c(\omega_l)$   relative to $T_c(\omega_l)$ , and also a noticeable shift of the dip in the dependence $h_c(\omega_l)$ . To explain this asymmetry, let us look at Fig.~\ref{fig15} showing the bisphere absorption spectra before and after collapse and the laser radiation spectrum.

For a negative initial shift $\Delta \omega = \omega _{l}-\omega _{01}<0$, the distance  $h$ is getting shorter under heating and the resonant frequency    $\omega _{01}$ \ ($\omega _{01} \rightarrow
\omega _1$) reduces either. As a result, the ($-\Delta \omega$) shift as well as the absorption grow and the heating increases.  That is, there is a positive feed-back between the radiation and the bisphere.
However if the initial  $\Delta
\omega$, the shift of   under heating leads to a lower absorption (a negative feed-back) and hence to a slower rate of collapse.

 So in the case of negative initial resonance detunings, $\Delta
\omega<0$, particles are heated to higher temperatures as compared to positive detunings, which accounts for the asymmetry of the temperature curves and the shift of the dip in the absorption spectrum during laser modification of large fractal aggregates.
The change of  $h_1$  as well as of  $\Delta\omega$ is negligible for larger initial distances $h$   (Fig.~\ref{fig13}) due the Van-der-Waals forces being small. This explains why photomodification depends so little on the sign of the shift.

\section{\label{part5}Conclusion}
As can be seen from the above discussion, the interaction of radiation with an elementary nanoparticle aggregate (a bisphere) involves a variety of processes. A detailed analysis of each of them is an independent, separate task. It is obvious that a comprehensive analysis of these problems is only possible on the basis of an interdisciplinary approach. Despite certain restrictions inherent in our model, it enabled us to establish a number of fundamental features of the photomodification process in a bisphere as well in a fractal aggregate through the change of its resonant domain structure. In particular, the results obtained prove that photomodification of fractal aggregates due to the radiation interaction with the lowest-frequency resonances of the long-wavelength band of the plasmon absorption of an aggregate (corresponding to small initial interparticle gaps) is caused by shortening of the interparticle distance and by the associated shift of resonances to the long-wavelength range. In the case of large initial values of interparticle gaps (thick adlayers) and radiation interaction with mainly high-frequency resonances of the long-wavelength absorption range of a fractal aggregate (being closer to $\omega_{pl}$ of a single particle), relative shifts of particles in a domain are insignificant even at high temperatures. Therefore the primary change in the domain has to be caused by the change in the size of particles (evaporation) down to their disappearance, and possibly by formation of plasma. As follows from the above reasoning, the nature of the photomodification energy threshold both for a bisphere and for nanoparticle aggregates in various spectral ranges can be different: at small interparticle gaps it is the shift of particles whereas at bigger interparticle gaps it is the change of the sizes of particles and their disappearance. Another important feature established in our studies at small initial values of interparticle gaps is the asymmetry of the spectral range of bisphere collapse with respect to the resonant frequency of the bisphere (the radiation frequency causing the collapse). As follows from Fig.~\ref{fig14}, the bisphere collapse is mainly observed when radiation is absorbed in the long-wavelength range of the bisphere absorption curve (Curve 1 in Fig.~\ref{fig15}). If a fractal aggregate is treated as a set of resonant domains (in the elementary case - bispheres) with various eigen frequencies, then collapse of bispheres will predominantly be observed for bispheres with frequencies exceeding $\omega_{l}$. In that case a dip will appear in the inhomogeneously broadened absorption spectrum of the fractal aggregate due to the laser-induced photomodification and the center of the dip will be somewhat shifted to the short-wavelength range from  $\omega_{l}$!

This result of the present study is of a primary importance, because the revealed asymmetry of the dip position, developing wihin the laser pulse duration, can be one of the main factors accounting for the negative nonlinear refraction of silver hydrosols observed in particular, in \cite{QE01,Ganeev} at the applied radiation wavelength  $\lambda_{l}=
1.064\ \mu$m. A detailed analysis of this result for nonlinear refraction and nonlinear absorption will be made elsewhere.
However it is obvious that such spectral changes become one of causes responsible for nonlinear absorption.

In the case of a short picosecond laser pulse, the approach of particles turns out to be much smaller, or absent at all (for example, at a low intensity which is close to the threshold values). Thus, our analysis shows that under conditions similar to the experimental ones the optical characteristics of the resonant domain of a fractal aggregate are dynamically changing within laser pulse duration, which generates nonlinearity of the optical response of the system and is manifested in various nonlinear-optical processes in pulsed laser fields described in the Introduction.

Our model, although a simplified one, enabled us to study the role of major factors and to estimate the necessity of taking them into account. For example, with the particles reaching high temperatures even at moderate intensities one has to take into consideration their evaporation. High temperatures of electrons make it obvious that their thermoemission should also be taken into account. As for the polymeric adlayer properties, the results obtained make it necessary to allow for the adlayer heterogeneity (the elasticity module of the adlayer increases in the direction of the particle surface). We could also mention a number of other problems we faced when using the developed model. In particular, evaporation leads to a decreased a size of particles and a reduced interparticle gap. The optical properties of a bisphere change as well as. Therefore the problem has to be considered for a bisphere with particles of a different size.

Besides, due to the principles incorporated in the model, it is possible to analyze the behavior of aggregates with not only metal particles but aslo with nanoparticles of any other material in a laser radiation field. It is quite obvious that the proposed model requires and what is important allows a further improvement and development to provide a more adequate description of the process of local modification and the associated wide variety of nonlinear optical phenomena. We emphasize that the processes observed in our study have an inevitable effect on all nonlinear-optical and photophysical phenomena which have been investigated so far but the physical nature of which remained obscure. One cannot rule out a possibility of similar processes occurring in nanocomposite materials containing metal nanoparticle aggregates in solid matrixes. The movement of particles in resonant domains in this case may become possible because of heating of particles by pulsed laser radiation and melting of the adjacent volume of the dehydrated polymer or even of the adjacent glass. Other processes though associated with, for example, particles evaporation can also occur in nanocomposites, resulting in anisotropy of the shape of the particles. The latter as well can affect the form and the dynamic shift of the plasmon absorption band \cite{Seifert}.
{\bf Acknowledgment}

Authors are thankful to prof. V.A. Markel (University of Pennsylvania) for calculations of polarizability of silver nanoparticle bispheres.
This research was supported by grants: DSP 2.1.1.1814, SS RAS 6612.2006.3, SB RAS 33.
\bibliography{Gavriluk_Karpov}

\begin{thebibliography}{65}
\expandafter\ifx\csname natexlab\endcsname\relax\def\natexlab#1{#1}\fi
\expandafter\ifx\csname bibnamefont\endcsname\relax
  \def\bibnamefont#1{#1}\fi
\expandafter\ifx\csname bibfnamefont\endcsname\relax
  \def\bibfnamefont#1{#1}\fi
\expandafter\ifx\csname citenamefont\endcsname\relax
  \def\citenamefont#1{#1}\fi
\expandafter\ifx\csname url\endcsname\relax
  \def\url#1{\texttt{#1}}\fi
\expandafter\ifx\csname urlprefix\endcsname\relax\def\urlprefix{URL }\fi
\providecommand{\bibinfo}[2]{#2}
\providecommand{\eprint}[2][]{\url{#2}}

\bibitem[{\citenamefont{Shalaev}(1996)}]{ShalPRep}
\bibinfo{author}{\bibfnamefont{V.~M.} \bibnamefont{Shalaev}},
  \bibinfo{journal}{Physics Reports} \textbf{\bibinfo{volume}{272}},
  \bibinfo{pages}{61} (\bibinfo{year}{1996}).

\bibitem[{\citenamefont{Shalaev}(2000)}]{Shalbook}
\bibinfo{author}{\bibfnamefont{V.~M.} \bibnamefont{Shalaev}},
  \emph{\bibinfo{title}{Nonlinear Optics of Random Media: Fractal Composites
  and Metal-Dielectric Films}} (\bibinfo{publisher}{Springer Verlag},
  \bibinfo{address}{Berlin}, \bibinfo{year}{2000}).

\bibitem[{\citenamefont{Karpov and Slabko}(2003)}]{bookKarpov}
\bibinfo{author}{\bibfnamefont{S.~V.} \bibnamefont{Karpov}} \bibnamefont{and}
  \bibinfo{author}{\bibfnamefont{V.~V.} \bibnamefont{Slabko}},
  \emph{\bibinfo{title}{Optical and Photophysical Properties of
  Fractal-Structured Metal Sols}} (\bibinfo{publisher}{Russian Academy of
  Sciences, Siberian Branch}, \bibinfo{address}{Novosibirsk},
  \bibinfo{year}{2003}).

\bibitem[{\citenamefont{Stockman et~al.}(1998)\citenamefont{Stockman, Pandey,
  and George}}]{stockman}
\bibinfo{author}{\bibfnamefont{M.~I.} \bibnamefont{Stockman}},
  \bibinfo{author}{\bibfnamefont{L.~N.} \bibnamefont{Pandey}},
  \bibnamefont{and} \bibinfo{author}{\bibfnamefont{T.~F.}
  \bibnamefont{George}}, \emph{\bibinfo{title}{Enhanced Nonlinear-Optical
  Responses of Disordered Clusters and Composites, in Nonlinear Optical
  Materials}} (\bibinfo{publisher}{Springer-Verlag}, \bibinfo{address}{New
  York}, \bibinfo{year}{1998}).

\bibitem[{\citenamefont{Shalaev et~al.}(1988)\citenamefont{Shalaev, Stockman,
  and George}}]{ZP88}
\bibinfo{author}{\bibfnamefont{V.~M.} \bibnamefont{Shalaev}},
  \bibinfo{author}{\bibfnamefont{M.~I.} \bibnamefont{Stockman}},
  \bibnamefont{and} \bibinfo{author}{\bibfnamefont{T.~F.}
  \bibnamefont{George}}, \bibinfo{journal}{Z. Phys. D}
  \textbf{\bibinfo{volume}{10}}, \bibinfo{pages}{71} (\bibinfo{year}{1988}).

\bibitem[{\citenamefont{Stockman et~al.}(1992)\citenamefont{Stockman, Shalaev,
  Moskovits, Botet, and George}}]{stockmanMarkel}
\bibinfo{author}{\bibfnamefont{M.~I.} \bibnamefont{Stockman}},
  \bibinfo{author}{\bibfnamefont{V.~M.} \bibnamefont{Shalaev}},
  \bibinfo{author}{\bibfnamefont{M.}~\bibnamefont{Moskovits}},
  \bibinfo{author}{\bibfnamefont{R.}~\bibnamefont{Botet}}, \bibnamefont{and}
  \bibinfo{author}{\bibfnamefont{T.~F.} \bibnamefont{George}},
  \bibinfo{journal}{Phys. Rev. B} \textbf{\bibinfo{volume}{46}},
  \bibinfo{pages}{2821} (\bibinfo{year}{1992}).

\bibitem[{\citenamefont{Danilova et~al.}(1993)\citenamefont{Danilova, Markel,
  and Safonov}}]{OAO}
\bibinfo{author}{\bibfnamefont{Y.~E.} \bibnamefont{Danilova}},
  \bibinfo{author}{\bibfnamefont{V.~A.} \bibnamefont{Markel}},
  \bibnamefont{and} \bibinfo{author}{\bibfnamefont{V.~P.}
  \bibnamefont{Safonov}}, \bibinfo{journal}{Atmos. Oceanic Opt.}
  \textbf{\bibinfo{volume}{6}}, \bibinfo{pages}{821} (\bibinfo{year}{1993}).

\bibitem[{\citenamefont{Shalaev et~al.}(1996)\citenamefont{Shalaev, Poliakov,
  and Markel}}]{stPOlMarkel}
\bibinfo{author}{\bibfnamefont{V.~M.} \bibnamefont{Shalaev}},
  \bibinfo{author}{\bibfnamefont{E.~Y.} \bibnamefont{Poliakov}},
  \bibnamefont{and} \bibinfo{author}{\bibfnamefont{V.~A.}
  \bibnamefont{Markel}}, \bibinfo{journal}{Phys. Rev. B}
  \textbf{\bibinfo{volume}{53}}, \bibinfo{pages}{2437} (\bibinfo{year}{1996}).

\bibitem[{\citenamefont{Butenko et~al.}(1990)\citenamefont{Butenko, Chubakov,
  Danilova, Karpov, Popov, Rautian, Safonov, Slabko, Shalaev, and
  Stockman}}]{but}
\bibinfo{author}{\bibfnamefont{A.~V.} \bibnamefont{Butenko}},
  \bibinfo{author}{\bibfnamefont{P.~A.} \bibnamefont{Chubakov}},
  \bibinfo{author}{\bibfnamefont{Y.~E.} \bibnamefont{Danilova}},
  \bibinfo{author}{\bibfnamefont{S.~V.} \bibnamefont{Karpov}},
  \bibinfo{author}{\bibfnamefont{A.~K.} \bibnamefont{Popov}},
  \bibinfo{author}{\bibfnamefont{S.~G.} \bibnamefont{Rautian}},
  \bibinfo{author}{\bibfnamefont{V.~P.} \bibnamefont{Safonov}},
  \bibinfo{author}{\bibfnamefont{V.~V.} \bibnamefont{Slabko}},
  \bibinfo{author}{\bibfnamefont{V.~M.} \bibnamefont{Shalaev}},
  \bibnamefont{and} \bibinfo{author}{\bibfnamefont{M.~I.}
  \bibnamefont{Stockman}}, \bibinfo{journal}{Phys. D}
  \textbf{\bibinfo{volume}{17}}, \bibinfo{pages}{283} (\bibinfo{year}{1990}).

\bibitem[{\citenamefont{Drachev et~al.}(2002)\citenamefont{Drachev, Perminov,
  Rautian, and Safonov}}]{Drach}
\bibinfo{author}{\bibfnamefont{V.~P.} \bibnamefont{Drachev}},
  \bibinfo{author}{\bibfnamefont{S.~V.} \bibnamefont{Perminov}},
  \bibinfo{author}{\bibfnamefont{S.~G.} \bibnamefont{Rautian}},
  \bibnamefont{and} \bibinfo{author}{\bibfnamefont{V.~P.}
  \bibnamefont{Safonov}}, \bibinfo{journal}{JETP}
  \textbf{\bibinfo{volume}{94}}, \bibinfo{pages}{901} (\bibinfo{year}{2002}).

\bibitem[{\citenamefont{Danilova et~al.}(1992)\citenamefont{Danilova,
  Plekhanov, and Safonov}}]{93}
\bibinfo{author}{\bibfnamefont{Y.~E.} \bibnamefont{Danilova}},
  \bibinfo{author}{\bibfnamefont{A.~I.} \bibnamefont{Plekhanov}},
  \bibnamefont{and} \bibinfo{author}{\bibfnamefont{V.~P.}
  \bibnamefont{Safonov}}, \bibinfo{journal}{Physica A}
  \textbf{\bibinfo{volume}{185}}, \bibinfo{pages}{61} (\bibinfo{year}{1992}).

\bibitem[{\citenamefont{Lepeshkin et~al.}(1999)\citenamefont{Lepeshkin, Kim,
  Safonov, Zhu, Armstrong, White, Zhur, and Shalaev}}]{Lepeshkin}
\bibinfo{author}{\bibfnamefont{N.~N.} \bibnamefont{Lepeshkin}},
  \bibinfo{author}{\bibfnamefont{W.}~\bibnamefont{Kim}},
  \bibinfo{author}{\bibfnamefont{V.~P.} \bibnamefont{Safonov}},
  \bibinfo{author}{\bibfnamefont{J.~G.} \bibnamefont{Zhu}},
  \bibinfo{author}{\bibfnamefont{R.~L.} \bibnamefont{Armstrong}},
  \bibinfo{author}{\bibfnamefont{C.~W.} \bibnamefont{White}},
  \bibinfo{author}{\bibfnamefont{R.~A.} \bibnamefont{Zhur}}, \bibnamefont{and}
  \bibinfo{author}{\bibfnamefont{V.~M.} \bibnamefont{Shalaev}},
  \bibinfo{journal}{Nonlinear Optical Physica and Materials}
  \textbf{\bibinfo{volume}{8}}, \bibinfo{pages}{191} (\bibinfo{year}{1999}).

\bibitem[{\citenamefont{Danilova et~al.}(1997)\citenamefont{Danilova,
  Lepeshkin, Rautian, and Safonov}}]{94}
\bibinfo{author}{\bibfnamefont{Y.~E.} \bibnamefont{Danilova}},
  \bibinfo{author}{\bibfnamefont{N.~N.} \bibnamefont{Lepeshkin}},
  \bibinfo{author}{\bibfnamefont{S.~G.} \bibnamefont{Rautian}},
  \bibnamefont{and} \bibinfo{author}{\bibfnamefont{V.~P.}
  \bibnamefont{Safonov}}, \bibinfo{journal}{Physica A}
  \textbf{\bibinfo{volume}{241}}, \bibinfo{pages}{231} (\bibinfo{year}{1997}).

\bibitem[{\citenamefont{Danilova
  et~al.}(1996{\natexlab{a}})\citenamefont{Danilova, Drachev, Perminov, and
  Safonov}}]{95}
\bibinfo{author}{\bibfnamefont{E.~Y.} \bibnamefont{Danilova}},
  \bibinfo{author}{\bibfnamefont{V.~P.} \bibnamefont{Drachev}},
  \bibinfo{author}{\bibfnamefont{S.~V.} \bibnamefont{Perminov}},
  \bibnamefont{and} \bibinfo{author}{\bibfnamefont{V.~P.}
  \bibnamefont{Safonov}}, \bibinfo{journal}{Bulletin of Russian Academy of
  Sci.} \textbf{\bibinfo{volume}{60}}, \bibinfo{pages}{18}
  (\bibinfo{year}{1996}{\natexlab{a}}).

\bibitem[{\citenamefont{Karpov et~al.}(2001)\citenamefont{Karpov, Kodirov,
  Ryasnyanskiy, and Slabko}}]{QE01}
\bibinfo{author}{\bibfnamefont{S.~V.} \bibnamefont{Karpov}},
  \bibinfo{author}{\bibfnamefont{M.~K.} \bibnamefont{Kodirov}},
  \bibinfo{author}{\bibfnamefont{A.~I.} \bibnamefont{Ryasnyanskiy}},
  \bibnamefont{and} \bibinfo{author}{\bibfnamefont{V.~V.}
  \bibnamefont{Slabko}}, \bibinfo{journal}{Quant. Electronics}
  \textbf{\bibinfo{volume}{31}}, \bibinfo{pages}{904} (\bibinfo{year}{2001}).

\bibitem[{\citenamefont{Ganeev et~al.}(2001)\citenamefont{Ganeev, Ryasnyansky,
  Kamalov, and Usmanov}}]{Ganeev}
\bibinfo{author}{\bibfnamefont{R.~A.} \bibnamefont{Ganeev}},
  \bibinfo{author}{\bibfnamefont{A.~I.} \bibnamefont{Ryasnyansky}},
  \bibinfo{author}{\bibfnamefont{S.~R.} \bibnamefont{Kamalov}},
  \bibnamefont{and} \bibinfo{author}{\bibfnamefont{T.~B.}
  \bibnamefont{Usmanov}}, \bibinfo{journal}{Phys. D: Appl. Phys.}
  \textbf{\bibinfo{volume}{34}}, \bibinfo{pages}{1602} (\bibinfo{year}{2001}).

\bibitem[{\citenamefont{Zhuravlev et~al.}(1992)\citenamefont{Zhuravlev, Orlova,
  Shelkovnikov, Plekhanov, Rautian, and Safonov}}]{Zhur}
\bibinfo{author}{\bibfnamefont{F.~A.} \bibnamefont{Zhuravlev}},
  \bibinfo{author}{\bibfnamefont{N.~A.} \bibnamefont{Orlova}},
  \bibinfo{author}{\bibfnamefont{V.~V.} \bibnamefont{Shelkovnikov}},
  \bibinfo{author}{\bibfnamefont{A.~I.} \bibnamefont{Plekhanov}},
  \bibinfo{author}{\bibfnamefont{S.~G.} \bibnamefont{Rautian}},
  \bibnamefont{and} \bibinfo{author}{\bibfnamefont{V.~P.}
  \bibnamefont{Safonov}}, \bibinfo{journal}{JETP Lett.}
  \textbf{\bibinfo{volume}{56}}, \bibinfo{pages}{264} (\bibinfo{year}{1992}).

\bibitem[{\citenamefont{Karpov et~al.}(1988)\citenamefont{Karpov, Popov,
  Rautian, Safonov, Slabko, Shalaev, and Shtokman}}]{JETP88}
\bibinfo{author}{\bibfnamefont{S.~V.} \bibnamefont{Karpov}},
  \bibinfo{author}{\bibfnamefont{A.~K.} \bibnamefont{Popov}},
  \bibinfo{author}{\bibfnamefont{S.~G.} \bibnamefont{Rautian}},
  \bibinfo{author}{\bibfnamefont{V.~P.} \bibnamefont{Safonov}},
  \bibinfo{author}{\bibfnamefont{V.~V.} \bibnamefont{Slabko}},
  \bibinfo{author}{\bibfnamefont{V.~M.} \bibnamefont{Shalaev}},
  \bibnamefont{and} \bibinfo{author}{\bibfnamefont{M.~I.}
  \bibnamefont{Shtokman}}, \bibinfo{journal}{JETP Lett.}
  \textbf{\bibinfo{volume}{48}}, \bibinfo{pages}{571} (\bibinfo{year}{1988}).

\bibitem[{\citenamefont{Safonov et~al.}(1998)\citenamefont{Safonov, Shalaev,
  Markel, Danilova, Lepeshkin, Kim, Rautian, and Armstrong}}]{PRL98}
\bibinfo{author}{\bibfnamefont{V.~P.} \bibnamefont{Safonov}},
  \bibinfo{author}{\bibfnamefont{V.~M.} \bibnamefont{Shalaev}},
  \bibinfo{author}{\bibfnamefont{V.~A.} \bibnamefont{Markel}},
  \bibinfo{author}{\bibfnamefont{Y.~E.} \bibnamefont{Danilova}},
  \bibinfo{author}{\bibfnamefont{N.~N.} \bibnamefont{Lepeshkin}},
  \bibinfo{author}{\bibfnamefont{W.}~\bibnamefont{Kim}},
  \bibinfo{author}{\bibfnamefont{S.~G.} \bibnamefont{Rautian}},
  \bibnamefont{and} \bibinfo{author}{\bibfnamefont{R.~L.}
  \bibnamefont{Armstrong}}, \bibinfo{journal}{Phys. Rev. Lett.}
  \textbf{\bibinfo{volume}{80}}, \bibinfo{pages}{1102} (\bibinfo{year}{1998}).

\bibitem[{\citenamefont{Karpov et~al.}(2003)\citenamefont{Karpov, Popov, and
  Slabko}}]{JTP03}
\bibinfo{author}{\bibfnamefont{S.~V.} \bibnamefont{Karpov}},
  \bibinfo{author}{\bibfnamefont{A.~K.} \bibnamefont{Popov}}, \bibnamefont{and}
  \bibinfo{author}{\bibfnamefont{V.~V.} \bibnamefont{Slabko}},
  \bibinfo{journal}{Technical Phys.} \textbf{\bibinfo{volume}{73}},
  \bibinfo{pages}{90} (\bibinfo{year}{2003}).

\bibitem[{\citenamefont{Karpov et~al.}(2002{\natexlab{a}})\citenamefont{Karpov,
  Bas'ko, Popov, and Slabko}}]{Ind}
\bibinfo{author}{\bibfnamefont{S.~V.} \bibnamefont{Karpov}},
  \bibinfo{author}{\bibfnamefont{A.~L.} \bibnamefont{Bas'ko}},
  \bibinfo{author}{\bibfnamefont{A.~K.} \bibnamefont{Popov}}, \bibnamefont{and}
  \bibinfo{author}{\bibfnamefont{V.~V.} \bibnamefont{Slabko}},
  \emph{\bibinfo{title}{Recent Research Developments in Optics}}
  (\bibinfo{publisher}{Research Signpost}, \bibinfo{address}{Trivandrum,
  India}, \bibinfo{year}{2002}{\natexlab{a}}).

\bibitem[{\citenamefont{Danilova
  et~al.}(1996{\natexlab{b}})\citenamefont{Danilova, Rautian, and
  Safonov}}]{Saf96}
\bibinfo{author}{\bibfnamefont{E.~Y.} \bibnamefont{Danilova}},
  \bibinfo{author}{\bibfnamefont{S.~G.} \bibnamefont{Rautian}},
  \bibnamefont{and} \bibinfo{author}{\bibfnamefont{V.~P.}
  \bibnamefont{Safonov}}, \bibinfo{journal}{Bulletin of Russian Academy of
  Sci.} \textbf{\bibinfo{volume}{60}}, \bibinfo{pages}{56}
  (\bibinfo{year}{1996}{\natexlab{b}}).

\bibitem[{\citenamefont{Kim et~al.}(1999)\citenamefont{Kim, Safonov, Shalaev,
  and Armstrong}}]{104}
\bibinfo{author}{\bibfnamefont{W.}~\bibnamefont{Kim}},
  \bibinfo{author}{\bibfnamefont{V.~P.} \bibnamefont{Safonov}},
  \bibinfo{author}{\bibfnamefont{V.~M.} \bibnamefont{Shalaev}},
  \bibnamefont{and} \bibinfo{author}{\bibfnamefont{R.~L.}
  \bibnamefont{Armstrong}}, \bibinfo{journal}{Phys. Rev. Lett.}
  \textbf{\bibinfo{volume}{82}}, \bibinfo{pages}{4811} (\bibinfo{year}{1999}).

\bibitem[{\citenamefont{Hache et~al.}(1988)\citenamefont{Hache, Richard,
  Flytzanis, and Kreibig}}]{Flyz}
\bibinfo{author}{\bibfnamefont{F.}~\bibnamefont{Hache}},
  \bibinfo{author}{\bibfnamefont{D.}~\bibnamefont{Richard}},
  \bibinfo{author}{\bibfnamefont{C.}~\bibnamefont{Flytzanis}},
  \bibnamefont{and} \bibinfo{author}{\bibfnamefont{U.}~\bibnamefont{Kreibig}},
  \bibinfo{journal}{Appl. Phys.} \textbf{\bibinfo{volume}{A47}},
  \bibinfo{pages}{347} (\bibinfo{year}{1988}).

\bibitem[{\citenamefont{Rautian}(1997)}]{Raut}
\bibinfo{author}{\bibfnamefont{S.~G.} \bibnamefont{Rautian}},
  \bibinfo{journal}{JETP} \textbf{\bibinfo{volume}{85}}, \bibinfo{pages}{451}
  (\bibinfo{year}{1997}).

\bibitem[{\citenamefont{Karpov et~al.}(2005)\citenamefont{Karpov, Gerasimov,
  Isaev, and Markel}}]{PRB05}
\bibinfo{author}{\bibfnamefont{S.~V.} \bibnamefont{Karpov}},
  \bibinfo{author}{\bibfnamefont{V.~S.} \bibnamefont{Gerasimov}},
  \bibinfo{author}{\bibfnamefont{I.~L.} \bibnamefont{Isaev}}, \bibnamefont{and}
  \bibinfo{author}{\bibfnamefont{V.~A.} \bibnamefont{Markel}},
  \bibinfo{journal}{Phys. Rev. B} \textbf{\bibinfo{volume}{72}},
  \bibinfo{pages}{205425} (\bibinfo{year}{2005}).

\bibitem[{\citenamefont{Karpov et~al.}(2006)\citenamefont{Karpov, Gerasimov,
  Isaev, and Markel}}]{JCP06}
\bibinfo{author}{\bibfnamefont{S.~V.} \bibnamefont{Karpov}},
  \bibinfo{author}{\bibfnamefont{V.~S.} \bibnamefont{Gerasimov}},
  \bibinfo{author}{\bibfnamefont{I.~L.} \bibnamefont{Isaev}}, \bibnamefont{and}
  \bibinfo{author}{\bibfnamefont{V.~A.} \bibnamefont{Markel}},
  \bibinfo{journal}{J. Chem. Phys.} \textbf{\bibinfo{volume}{125}},
  \bibinfo{pages}{111101} (\bibinfo{year}{2006}).

\bibitem[{\citenamefont{Karpov et~al.}(2007)\citenamefont{Karpov, Gerasimov,
  Isaev, Podavalova, and Slabko}}]{CJ07}
\bibinfo{author}{\bibfnamefont{S.~V.} \bibnamefont{Karpov}},
  \bibinfo{author}{\bibfnamefont{V.~S.} \bibnamefont{Gerasimov}},
  \bibinfo{author}{\bibfnamefont{I.~L.} \bibnamefont{Isaev}},
  \bibinfo{author}{\bibfnamefont{O.~P.} \bibnamefont{Podavalova}},
  \bibnamefont{and} \bibinfo{author}{\bibfnamefont{V.~V.}
  \bibnamefont{Slabko}}, \bibinfo{journal}{Colloid J.}
  \textbf{\bibinfo{volume}{69}}, \bibinfo{pages}{159} (\bibinfo{year}{2007}).

\bibitem[{\citenamefont{Claro}(1994)}]{Claro1}
\bibinfo{author}{\bibfnamefont{F.}~\bibnamefont{Claro}},
  \bibinfo{journal}{Appl. Phys. Lett.} \textbf{\bibinfo{volume}{65}},
  \bibinfo{pages}{2743} (\bibinfo{year}{1994}).

\bibitem[{\citenamefont{Claro}(1997)}]{Claro2}
\bibinfo{author}{\bibfnamefont{F.}~\bibnamefont{Claro}},
  \bibinfo{journal}{Physica A} \textbf{\bibinfo{volume}{241}},
  \bibinfo{pages}{223} (\bibinfo{year}{1997}).

\bibitem[{\citenamefont{Fuchs and Claro}(2004)}]{Claro3}
\bibinfo{author}{\bibfnamefont{R.}~\bibnamefont{Fuchs}} \bibnamefont{and}
  \bibinfo{author}{\bibfnamefont{F.}~\bibnamefont{Claro}},
  \bibinfo{journal}{Appl. Phys. Lett.} \textbf{\bibinfo{volume}{35}},
  \bibinfo{pages}{3280} (\bibinfo{year}{2004}).

\bibitem[{\citenamefont{Hallok et~al.}(2005)\citenamefont{Hallok, Redmond, and
  Brus}}]{Hallok}
\bibinfo{author}{\bibfnamefont{A.~J.} \bibnamefont{Hallok}},
  \bibinfo{author}{\bibfnamefont{P.~L.} \bibnamefont{Redmond}},
  \bibnamefont{and} \bibinfo{author}{\bibfnamefont{L.~E.} \bibnamefont{Brus}},
  \bibinfo{journal}{Proc. NAS} \textbf{\bibinfo{volume}{102}},
  \bibinfo{pages}{1280} (\bibinfo{year}{2005}).

\bibitem[{\citenamefont{Karpov et~al.}(2002{\natexlab{b}})\citenamefont{Karpov,
  Slabko, and Chiganova}}]{CJ2002}
\bibinfo{author}{\bibfnamefont{S.~V.} \bibnamefont{Karpov}},
  \bibinfo{author}{\bibfnamefont{V.~V.} \bibnamefont{Slabko}},
  \bibnamefont{and} \bibinfo{author}{\bibfnamefont{G.~A.}
  \bibnamefont{Chiganova}}, \bibinfo{journal}{Colloid J.}
  \textbf{\bibinfo{volume}{64}}, \bibinfo{pages}{474}
  (\bibinfo{year}{2002}{\natexlab{b}}).

\bibitem[{\citenamefont{Markel et~al.}(2004)\citenamefont{Markel, Pustovit,
  Karpov, Obuschenko, Gerasimov, and Isaev}}]{PRB04}
\bibinfo{author}{\bibfnamefont{V.~A.} \bibnamefont{Markel}},
  \bibinfo{author}{\bibfnamefont{V.~N.} \bibnamefont{Pustovit}},
  \bibinfo{author}{\bibfnamefont{S.~V.} \bibnamefont{Karpov}},
  \bibinfo{author}{\bibfnamefont{A.~V.} \bibnamefont{Obuschenko}},
  \bibinfo{author}{\bibfnamefont{V.~S.} \bibnamefont{Gerasimov}},
  \bibnamefont{and} \bibinfo{author}{\bibfnamefont{I.~L.} \bibnamefont{Isaev}},
  \bibinfo{journal}{Phys. Rev. B} \textbf{\bibinfo{volume}{70}},
  \bibinfo{pages}{054202} (\bibinfo{year}{2004}).

\bibitem[{\citenamefont{Heard et~al.}(1983)\citenamefont{Heard, Griezer,
  Barrachlough, and Sanders}}]{Heard}
\bibinfo{author}{\bibfnamefont{S.~M.} \bibnamefont{Heard}},
  \bibinfo{author}{\bibfnamefont{F.}~\bibnamefont{Griezer}},
  \bibinfo{author}{\bibfnamefont{C.~G.} \bibnamefont{Barrachlough}},
  \bibnamefont{and} \bibinfo{author}{\bibfnamefont{J.~V.}
  \bibnamefont{Sanders}}, \bibinfo{journal}{J. Colloid. Interface Sci.}
  \textbf{\bibinfo{volume}{93}}, \bibinfo{pages}{545} (\bibinfo{year}{1983}).

\bibitem[{\citenamefont{Sheik-Bahae et~al.}(1990)\citenamefont{Sheik-Bahae,
  Said, Wei, Hagan, and van Stryland}}]{Zscan}
\bibinfo{author}{\bibfnamefont{M.}~\bibnamefont{Sheik-Bahae}},
  \bibinfo{author}{\bibfnamefont{A.~A.} \bibnamefont{Said}},
  \bibinfo{author}{\bibfnamefont{T.~H.} \bibnamefont{Wei}},
  \bibinfo{author}{\bibfnamefont{D.}~\bibnamefont{Hagan}}, \bibnamefont{and}
  \bibinfo{author}{\bibfnamefont{E.~W.} \bibnamefont{van Stryland}},
  \bibinfo{journal}{IEEE J. Quantum Electron} \textbf{\bibinfo{volume}{26}},
  \bibinfo{pages}{760} (\bibinfo{year}{1990}).

\bibitem[{\citenamefont{Shen}(1984)}]{Shen}
\bibinfo{author}{\bibfnamefont{Y.~R.} \bibnamefont{Shen}},
  \emph{\bibinfo{title}{The principles of nonlinear optics}}
  (\bibinfo{publisher}{John Willey and sons, Inc}, \bibinfo{address}{New York},
  \bibinfo{year}{1984}).

\bibitem[{\citenamefont{Hale and Querry}(1973)}]{115}
\bibinfo{author}{\bibfnamefont{G.~M.} \bibnamefont{Hale}} \bibnamefont{and}
  \bibinfo{author}{\bibfnamefont{M.~R.} \bibnamefont{Querry}},
  \bibinfo{journal}{Appl. Optics.} \textbf{\bibinfo{volume}{12}},
  \bibinfo{pages}{555} (\bibinfo{year}{1973}).

\bibitem[{\citenamefont{Kerr et~al.}(1972)\citenamefont{Kerr, Hamm, Williams,
  Birkholf, and Painter}}]{116}
\bibinfo{author}{\bibfnamefont{G.~D.} \bibnamefont{Kerr}},
  \bibinfo{author}{\bibfnamefont{R.~N.} \bibnamefont{Hamm}},
  \bibinfo{author}{\bibfnamefont{M.~W.} \bibnamefont{Williams}},
  \bibinfo{author}{\bibfnamefont{R.~D.} \bibnamefont{Birkholf}},
  \bibnamefont{and} \bibinfo{author}{\bibfnamefont{L.~R.}
  \bibnamefont{Painter}}, \bibinfo{journal}{Phys. Rev. A.}
  \textbf{\bibinfo{volume}{5}}, \bibinfo{pages}{2523} (\bibinfo{year}{1972}).

\bibitem[{\citenamefont{Feng et~al.}(1995)\citenamefont{Feng, Moloney, Newell,
  and Wright}}]{13}
\bibinfo{author}{\bibfnamefont{Q.}~\bibnamefont{Feng}},
  \bibinfo{author}{\bibfnamefont{J.~V.} \bibnamefont{Moloney}},
  \bibinfo{author}{\bibfnamefont{A.~C.} \bibnamefont{Newell}},
  \bibnamefont{and} \bibinfo{author}{\bibfnamefont{E.~M.}
  \bibnamefont{Wright}}, \bibinfo{journal}{Optics Letters}
  \textbf{\bibinfo{volume}{20}}, \bibinfo{pages}{1958} (\bibinfo{year}{1995}).

\bibitem[{\citenamefont{Tsang}(1996)}]{14}
\bibinfo{author}{\bibfnamefont{T.}~\bibnamefont{Tsang}}, \bibinfo{journal}{Phys
  Rev A} \textbf{\bibinfo{volume}{54}}, \bibinfo{pages}{5454}
  (\bibinfo{year}{1996}).

\bibitem[{\citenamefont{Eliashevich}(1962)}]{Eljashevich}
\bibinfo{author}{\bibfnamefont{M.~A.} \bibnamefont{Eliashevich}},
  \emph{\bibinfo{title}{Spectroscopy of atoms and molecules}}
  (\bibinfo{publisher}{State Publ. House of Phys.-Math. Literature.},
  \bibinfo{address}{Moscow}, \bibinfo{year}{1962}).

\bibitem[{\citenamefont{Askar'yan}(1966)}]{16}
\bibinfo{author}{\bibfnamefont{G.~A.} \bibnamefont{Askar'yan}},
  \bibinfo{journal}{JETP Lett.} \textbf{\bibinfo{volume}{4}},
  \bibinfo{pages}{400} (\bibinfo{year}{1966}).

\bibitem[{17(1974)}]{17}
\emph{\bibinfo{title}{Energies of chemical bond opening. Ionisation potecials
  and affinity to electron. Handbook}} (\bibinfo{publisher}{Nauka},
  \bibinfo{address}{Moscow}, \bibinfo{year}{1974}).

\bibitem[{\citenamefont{Chastov and Lebedev}(1970)}]{Chastov}
\bibinfo{author}{\bibfnamefont{A.~A.} \bibnamefont{Chastov}} \bibnamefont{and}
  \bibinfo{author}{\bibfnamefont{O.~L.} \bibnamefont{Lebedev}},
  \bibinfo{journal}{JETP} \textbf{\bibinfo{volume}{58}}, \bibinfo{pages}{848}
  (\bibinfo{year}{1970}).

\bibitem[{\citenamefont{Bohren and Huffman}(1983)}]{Boren}
\bibinfo{author}{\bibfnamefont{C.~F.} \bibnamefont{Bohren}} \bibnamefont{and}
  \bibinfo{author}{\bibfnamefont{D.~R.} \bibnamefont{Huffman}},
  \emph{\bibinfo{title}{Absorption and Scattering of Light by Small Particles}}
  (\bibinfo{publisher}{Wiley}, \bibinfo{address}{New York},
  \bibinfo{year}{1983}).

\bibitem[{\citenamefont{Johnson and Christy}(1972)}]{Christi}
\bibinfo{author}{\bibfnamefont{P.~B.} \bibnamefont{Johnson}} \bibnamefont{and}
  \bibinfo{author}{\bibfnamefont{R.~W.} \bibnamefont{Christy}},
  \bibinfo{journal}{Phys. Rev. B.} \textbf{\bibinfo{volume}{6}},
  \bibinfo{pages}{4370} (\bibinfo{year}{1972}).

\bibitem[{\citenamefont{Markel et~al.}(1996)\citenamefont{Markel, Shalaev,
  Stechel, Kim, and Armstrong}}]{Markel}
\bibinfo{author}{\bibfnamefont{V.~A.} \bibnamefont{Markel}},
  \bibinfo{author}{\bibfnamefont{V.~M.} \bibnamefont{Shalaev}},
  \bibinfo{author}{\bibfnamefont{E.~B.} \bibnamefont{Stechel}},
  \bibinfo{author}{\bibfnamefont{W.}~\bibnamefont{Kim}}, \bibnamefont{and}
  \bibinfo{author}{\bibfnamefont{R.~L.} \bibnamefont{Armstrong}},
  \bibinfo{journal}{Phys. Rev. B.} \textbf{\bibinfo{volume}{53}},
  \bibinfo{pages}{2425} (\bibinfo{year}{1996}).

\bibitem[{\citenamefont{Nieto-Vesperinas
  et~al.}(2004)\citenamefont{Nieto-Vesperinas, Chaumet, and
  Rahmani}}]{Vesperinas}
\bibinfo{author}{\bibfnamefont{M.}~\bibnamefont{Nieto-Vesperinas}},
  \bibinfo{author}{\bibfnamefont{P.~C.} \bibnamefont{Chaumet}},
  \bibnamefont{and} \bibinfo{author}{\bibfnamefont{A.}~\bibnamefont{Rahmani}},
  \bibinfo{journal}{Phil. Trasn. R. Soc. Lond. A.}
  \textbf{\bibinfo{volume}{362}}, \bibinfo{pages}{719} (\bibinfo{year}{2004}).

\bibitem[{\citenamefont{Landau and Lifshits}(1999)}]{Landau_Upr}
\bibinfo{author}{\bibfnamefont{L.~D.} \bibnamefont{Landau}} \bibnamefont{and}
  \bibinfo{author}{\bibfnamefont{E.~M.} \bibnamefont{Lifshits}},
  \emph{\bibinfo{title}{The theory of elasticity. Course of Theoretical
  Physics}} (\bibinfo{publisher}{Butterworth-Heinemann},
  \bibinfo{address}{Oxford}, \bibinfo{year}{1999}).

\bibitem[{\citenamefont{Lewis}(2000)}]{Lewis}
\bibinfo{author}{\bibfnamefont{J.~A.} \bibnamefont{Lewis}},
  \bibinfo{journal}{J. Am. Ceram. Soc.} \textbf{\bibinfo{volume}{83}},
  \bibinfo{pages}{2341} (\bibinfo{year}{2000}).

\bibitem[{\citenamefont{Pankratova and Izmailova}(1976)}]{Izmailova}
\bibinfo{author}{\bibfnamefont{M.~N.} \bibnamefont{Pankratova}}
  \bibnamefont{and} \bibinfo{author}{\bibfnamefont{V.~N.}
  \bibnamefont{Izmailova}}, \bibinfo{journal}{Colloid J.}
  \textbf{\bibinfo{volume}{38}}, \bibinfo{pages}{490} (\bibinfo{year}{1976}).

\bibitem[{\citenamefont{Slutsker et~al.}(2002)\citenamefont{Slutsker,
  Polikarpov, and Vasilieva}}]{Slutsker}
\bibinfo{author}{\bibfnamefont{A.~I.} \bibnamefont{Slutsker}},
  \bibinfo{author}{\bibfnamefont{Y.~I.} \bibnamefont{Polikarpov}},
  \bibnamefont{and} \bibinfo{author}{\bibfnamefont{K.~F.}
  \bibnamefont{Vasilieva}}, \bibinfo{journal}{Solid State Physics}
  \textbf{\bibinfo{volume}{44}}, \bibinfo{pages}{1604} (\bibinfo{year}{2002}).

\bibitem[{\citenamefont{Frenkel}(1975)}]{Frenkel}
\bibinfo{author}{\bibfnamefont{Y.~A.} \bibnamefont{Frenkel}},
  \emph{\bibinfo{title}{The Kinetic theory of liquids}}
  (\bibinfo{publisher}{Nauka}, \bibinfo{address}{Moscow},
  \bibinfo{year}{1975}).

\bibitem[{\citenamefont{Pustovalov et~al.}(1988)\citenamefont{Pustovalov,
  Khorunzhyi, and Bobuchenko}}]{Pustovalov}
\bibinfo{author}{\bibfnamefont{V.~K.} \bibnamefont{Pustovalov}},
  \bibinfo{author}{\bibfnamefont{I.~A.} \bibnamefont{Khorunzhyi}},
  \bibnamefont{and} \bibinfo{author}{\bibfnamefont{D.~S.}
  \bibnamefont{Bobuchenko}}, \bibinfo{journal}{Bulletin of Academy of Sci. of
  USSR} \textbf{\bibinfo{volume}{52}}, \bibinfo{pages}{1847}
  (\bibinfo{year}{1988}).

\bibitem[{\citenamefont{Frolov}(1982)}]{Frolov}
\bibinfo{author}{\bibfnamefont{Y.~G.} \bibnamefont{Frolov}},
  \emph{\bibinfo{title}{Textbook of Colloid Chemistry}}
  (\bibinfo{publisher}{Khimia}, \bibinfo{address}{Moscow},
  \bibinfo{year}{1982}).

\bibitem[{\citenamefont{Sonntag and Strenge}(1970)}]{Zontag}
\bibinfo{author}{\bibfnamefont{H.}~\bibnamefont{Sonntag}} \bibnamefont{and}
  \bibinfo{author}{\bibfnamefont{K.}~\bibnamefont{Strenge}},
  \emph{\bibinfo{title}{Koagulation und stabilitat disperser systeme}}
  (\bibinfo{publisher}{VEB Deucher Verlag Der Wissenschaften},
  \bibinfo{address}{Berlin}, \bibinfo{year}{1970}).

\bibitem[{\citenamefont{Ansell and Dickinson}(1987)}]{Ansell}
\bibinfo{author}{\bibfnamefont{G.~C.} \bibnamefont{Ansell}} \bibnamefont{and}
  \bibinfo{author}{\bibfnamefont{E.}~\bibnamefont{Dickinson}},
  \bibinfo{journal}{Phys. Rev. A.} \textbf{\bibinfo{volume}{35}},
  \bibinfo{pages}{2349} (\bibinfo{year}{1987}).

\bibitem[{\citenamefont{Enustun and J. Turkevich}(1963)}]{Enustun}
\bibinfo{author}{\bibfnamefont{B.}~\bibnamefont{Enustun}} \bibnamefont{and}
  \bibinfo{author}{\bibnamefont{J. Turkevich}}, \bibinfo{journal}{J. Am. Chem.
  Soc.} \textbf{\bibinfo{volume}{85}}, \bibinfo{pages}{3317}
  (\bibinfo{year}{1963}).

\bibitem[{\citenamefont{Sauer and Lowen}(1996)}]{Sauer}
\bibinfo{author}{\bibfnamefont{S.}~\bibnamefont{Sauer}} \bibnamefont{and}
  \bibinfo{author}{\bibfnamefont{H.}~\bibnamefont{Lowen}}, \bibinfo{journal}{J.
  Phys.: Condens. Matter.} \textbf{\bibinfo{volume}{8}}, \bibinfo{pages}{L803}
  (\bibinfo{year}{1996}).

\bibitem[{\citenamefont{Deriaguin}(1986)}]{Derjagin}
\bibinfo{author}{\bibfnamefont{B.~V.} \bibnamefont{Deriaguin}},
  \emph{\bibinfo{title}{Theory of stability of colloids and thin films}}
  (\bibinfo{publisher}{Nauka}, \bibinfo{address}{Moscow},
  \bibinfo{year}{1986}).

\bibitem[{\citenamefont{Anisimov et~al.}(1970)\citenamefont{Anisimov, Imas,
  Romanov, and Hodyko}}]{Anisimov}
\bibinfo{author}{\bibfnamefont{S.~I.} \bibnamefont{Anisimov}},
  \bibinfo{author}{\bibfnamefont{Y.~A.} \bibnamefont{Imas}},
  \bibinfo{author}{\bibfnamefont{G.~S.} \bibnamefont{Romanov}},
  \bibnamefont{and} \bibinfo{author}{\bibfnamefont{Y.~V.}
  \bibnamefont{Hodyko}}, \emph{\bibinfo{title}{The effects of high-power laser
  radiation on metals}} (\bibinfo{publisher}{Nauka}, \bibinfo{address}{Moscow},
  \bibinfo{year}{1970}).

\bibitem[{\citenamefont{B.Wright}(1994)}]{wright}
\bibinfo{author}{\bibfnamefont{O.}~\bibnamefont{B.Wright}},
  \bibinfo{journal}{Phys. Rev. B.} \textbf{\bibinfo{volume}{49}},
  \bibinfo{pages}{9985} (\bibinfo{year}{1994}).

\bibitem[{\citenamefont{Belousova et~al.}(2003)\citenamefont{Belousova,
  Mironova, and Yur'ev}}]{Belousova}
\bibinfo{author}{\bibfnamefont{I.~M.} \bibnamefont{Belousova}},
  \bibinfo{author}{\bibfnamefont{N.~G.} \bibnamefont{Mironova}},
  \bibnamefont{and} \bibinfo{author}{\bibfnamefont{N.~S.}
  \bibnamefont{Yur'ev}}, \bibinfo{journal}{Optics and Spectroscopy}
  \textbf{\bibinfo{volume}{94}}, \bibinfo{pages}{93} (\bibinfo{year}{2003}).

\bibitem[{\citenamefont{Seifert et~al.}(2001)\citenamefont{Seifert, Kaempfe,
  Berg, and Graener}}]{Seifert}
\bibinfo{author}{\bibfnamefont{G.}~\bibnamefont{Seifert}},
  \bibinfo{author}{\bibfnamefont{M.}~\bibnamefont{Kaempfe}},
  \bibinfo{author}{\bibfnamefont{K.-J.} \bibnamefont{Berg}}, \bibnamefont{and}
  \bibinfo{author}{\bibfnamefont{H.}~\bibnamefont{Graener}},
  \bibinfo{journal}{Appl. Phys. B.} \textbf{\bibinfo{volume}{73}},
  \bibinfo{pages}{355} (\bibinfo{year}{2001}).

\end{thebibliography}

\end{document}